\documentclass[
reprint,
superscriptaddress,
aps,
pre,
longbibliography,
showpacs
]{revtex4-1}

\usepackage{graphicx}
\usepackage{dcolumn}
\usepackage{bm}
\usepackage[utf8]{inputenc}
\usepackage{amsmath, amsthm, amssymb, amsfonts}
\usepackage{accents}
\usepackage[usenames,dvipsnames]{xcolor}
\usepackage[caption=false]{subfig}
\usepackage{tikz}
\usepackage{array}

\usetikzlibrary{shapes.geometric}
\usetikzlibrary{shapes.misc}
\usetikzlibrary{arrows.meta}

\newcolumntype{M}[1]{>{\centering\arraybackslash}m{#1}}

\newcommand{\YZ}{\color{black}}
\newcommand{\TN}{\color{black}}

\newcommand{\Fi}{F^{(i)}}

\begin{document}

\newcommand{\mytitle}{Asymmetry-Induced Synchronization in Oscillator Networks}
\title{\mytitle}

\author{Yuanzhao Zhang}
\affiliation{Department of Physics and Astronomy, Northwestern University, Evanston, Illinois 60208, USA}
\author{Takashi Nishikawa}
\author{Adilson E. Motter}
\affiliation{Department of Physics and Astronomy, Northwestern University, Evanston, Illinois 60208, USA}
\affiliation{Northwestern Institute on Complex Systems, Northwestern University, Evanston, Illinois 60208, USA}

\begin{abstract}
A scenario has recently been reported in which in order to stabilize complete synchronization of an oscillator network---a symmetric state---the symmetry of the system itself has to be broken by making the oscillators nonidentical. 
But how often does such behavior---which we term asymmetry-induced synchronization (\textit{AISync})---occur in oscillator networks? 
Here we present the first general scheme for constructing \textit{AISync} systems and demonstrate that this behavior is the norm rather than the exception in a wide class of physical systems that can be seen as multilayer networks. 
Since a symmetric network in complete synchrony is the basic building block of cluster synchronization in more general networks, \textit{AISync} should be common also in facilitating cluster synchronization by breaking the symmetry of the cluster subnetworks.
\end{abstract}

\pacs{05.45.Xt, 89.75.Fb}

\maketitle

\section{Introduction}

A common assumption in the field of network dynamics is that homogeneity in the local dynamics~\cite{restrepo2004spatial,sun2009master} and interaction network~\cite{nishikawa2003heterogeneity,denker2004breaking,donetti2005entangled}---or in the combination of both~\cite{motter2005network,zhou2006dynamical}---can facilitate complete synchronization.
It has been recently shown, however, that structural heterogeneity in networks of identical oscillators~\cite{nishikawa2010network} or oscillator heterogeneity in structurally symmetric networks~\cite{PhysRevLett.117.114101} can stabilize otherwise unstable synchronous states, thus effectively breaking the symmetry of a system to stabilize a symmetric state.
These scenarios, which we refer to as \textit{asymmetry-induced synchronization} ({\it AISync}), can be interpreted as the converse of symmetry breaking, and hence as a converse of chimera states~\cite{kuramoto2002coexistence,abrams2004chimera}. 
Perhaps the most striking and the strongest form of {\it AISync} is the one in which oscillators coupled in a symmetric network (i.e., each oscillator plays exactly the same structural role) can converge to identical dynamics only when they themselves are nonidentical; this has been demonstrated, however, exclusively for rotationally symmetric networks and one type of periodic oscillators~\cite{PhysRevLett.117.114101}.
Whether such {\it AISync} behavior can be shown to be common among systems with other symmetric network structures and oscillator dynamics, including experimentally testable ones, has been an open question.

In this article we introduce and analyze a broad class of {\it AISync} systems that can have general symmetric network structure with multiple link types and general oscillator dynamics (which can be chaotic, periodic, continuous-time, discrete-time, etc.). This in particular includes physical systems previously used in network synchronization experiments, thus providing a recipe for future empirical studies.
For this class, we demonstrate that {\it AISync} is indeed common and provide a full characterization of those networks that support {\it AISync} behavior, showing that the fraction of such networks is significant over a range of network sizes and link densities.


\section{Definition of \textit{AIS\MakeLowercase{ync}}}

\begin{figure*}[hbt!]
\includegraphics[width=1\textwidth]{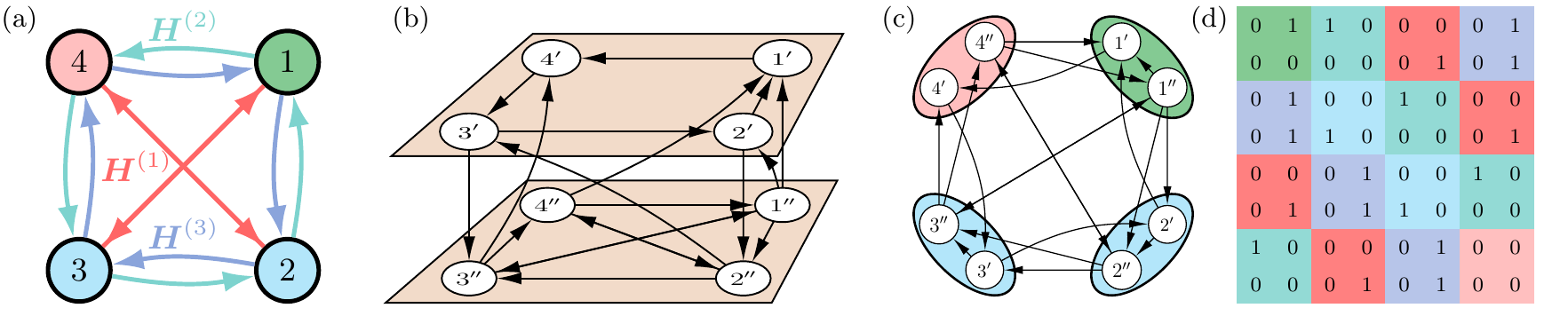}
\caption{
Multilayer construction of \textit{AISync} networks.
(a) Example of a symmetric network of $N=4$ heterogeneous oscillators and $K=3$ types of (directed) links with associated coupling functions $\bm{H}^{(1)}$, $\bm{H}^{(2)}$ and $\bm{H}^{(3)}$.
(b) One of many possible multilayer networks corresponding to the network in (a), with $L=2$ layers and $n=LN=8$ identical subnodes.
Subnodes are labeled with node indices, with prime and double prime indicating layer $1$ and $2$, respectively.
(c) Flattened, monolayer representation of the multilayer network in (b).
(d) Block structure of the adjacency matrix $\widetilde{A}$ for the monolayer network in (c).
Colors indicate different types of nodes ($F^{(i)}$, diagonal blocks) and links ($\widetilde{A}^{(ii’)}$, off-diagonal blocks). }
\label{fig:1}
\end{figure*}

To formulate a precise definition of \textit{AISync}, we consider networks of $N$ (not necessarily identical) oscillators coupled through $K$ different types of interactions. The network dynamics is described by
\begin{equation}
\dot{\bm{X}}_i = \bm{F}_i(\bm{X}_i) + \sum_{\alpha=1}^K \sum_{\substack{i'=1\\ i'\neq i}}^N A^{(\alpha)}_{ii'} \bm{H}^{(\alpha)}(\bm{X}_i, \bm{X}_{i'}),
\label{eq:0}
\end{equation}
where $\bm{X}_i = \bm{X}_i(t)$ is the $M$-dimensional state vector of node $i$, the function $\bm{F}_i$ governs the intrinsic dynamics of node $i$, the adjacency matrix $A^{(\alpha)}=(A^{(\alpha)}_{ii'})$ represents the topology of interactions through links of type $\alpha$, and $\bm{H}^{(\alpha)}$ is the coupling function associated with the link type $\alpha$.
A completely synchronous state of the network is defined by $\bm{X}_1(t)=\bm{X}_2(t)=\cdots=\bm{X}_N(t)$.

To isolate the effect of breaking the homogeneity of oscillators, we consider adjacency matrices $A^{(\alpha)}$ that together represent a {\it symmetric network}, defined as a network in which every node can be mapped to any other node by some permutation of nodes without changing any $A^{(\alpha)}$.
Thus, the set of links of any given type must couple every node identically (see Fig.~\ref{fig:1}(a) for an example).
{\TN
The rationale for using symmetric network structures here is to ensure that any stabilization of complete synchronization by oscillator heterogeneity is due to the reduced system symmetry (as required for {\it AISync}) and not due to having network heterogeneity and oscillator heterogeneity compensating each other, which may not break the system symmetry.
}

When restricted to undirected networks with a single link type, our definition of symmetric networks yields the class of vertex-transitive graphs from graph theory~\cite{biggs1993algebraic}.
This rich class encompasses Cayley graphs (defined as a network of relations between elements of a finite group; Appendix~\ref{sec:cayley}) and circulant graphs (defined as a network whose nodes can be arranged in a ring so that the network is invariant under rotations), which have previously been used to study chimera states~\cite{panaggio2015chimera}.
Enumerating {\it all} vertex-transitive graphs of a given size $N$ becomes challenging as $N$ grows and has so far been completed only for $N < 32$~\cite{oeis}.
The symmetric networks we consider here generalizes vertex-transitive graphs to the even richer class of networks that can be directed and include multiple link types.

Given a symmetric network structure, the system in Eq.~\eqref{eq:0} exhibits {\it AISync} if it satisfies the following two conditions:
(C1) there are no asymptotically stable synchronous states for any {\it homogeneous system} (i.e., with $\bm{F}_1=\cdots=\bm{F}_N$), and
(C2) there is an asymptotically {\it heterogeneous system} (i.e., with $\bm{F}_i \neq \bm{F}_{i'}$, for some $i \neq i'$) for which a stable synchronous state exists.
A challenge in establishing {\it AISync} is that the form of Eq.~\eqref{eq:0} does not guarantee the existence of a completely synchronous state.
Another challenge concerns the stability analysis of such a state, since Eq.~\eqref{eq:0} is beyond the framework normally used in the master stability function (MSF) approach and its generalizations currently available~\cite{sun2009master,dahms2012cluster,pecora2014cluster,Genioe1601679}: oscillators can be nonidentical (different $\bm{F}_i$), and the network can host $K>1$ types of directed interactions.

\section{Multilayer systems considered}

To overcome these challenges, below we propose a {\it multilayer construction} that defines a large, general subclass of systems within the class given by Eq.~\eqref{eq:0}.  
We show that any system in this subclass is guaranteed to have a synchronous state, and the stability of that state can be analyzed by applying the MSF framework to the flattened representation of the system.
{\TN 
The MSF approach decouples the oscillator dynamics from the network structure, which enables us to draw conclusions about {\it AISync} for general oscillator dynamics.
}

In our multilayer system, each node is composed of $L$ identical {\it subnodes}, belonging to $L$ different layers and connected by a set of {\it internal sublinks}.
The pattern of these internal sublinks is thus part of the node's properties and determines the heterogeneity across nodes.
For a pair of connected nodes, the type of the connecting link is determined by the pattern of {\it external sublinks} between the subnodes of these two nodes.
This construction yields a multilayer network~\cite{gao2012networks,gomez2013diffusion,de2013mathematical,boccaletti2014structure,de2016physics} of subnodes and sublinks with $L$ layers; 
see Fig.~\ref{fig:1}(b) for an $L=2$ example.
Note that in general there is more than one possible multilayer network for a given symmetric network.
Networks with such layered structure have been used extensively as realistic models of various natural and man-made systems. The class of systems just defined is broader than most classes of systems used in previous studies of synchronization on multilayer networks~\cite{zhang2015explosive,gambuzza2015intra}, since the links between two different layers are not constrained to be one-to-one. 
The underlying hierarchical organization, in which each node decomposes into interacting subnodes, is shared by many physical systems, such as the multi-processor nodes in modern supercomputers.

Coupling the dynamics of subnodes diffusively in this network, the multilayer system can be described at the subnode level as
\begin{equation}
\dot{\bm{x}}^{(i)}_{\ell}
= \bm{f}(\bm{x}^{(i)}_{\ell}) 
+ \sum_{i'=1}^N \sum_{\ell'=1}^L \widetilde{A}^{(ii')}_{\ell\ell'} \bigl[\bm{h}(\bm{x}^{(i')}_{\ell'}) - \bm{h}(\bm{x}^{(i)}_{\ell}) \bigr],
\label{eq:multilayer}
\end{equation}
where $\bm{x}^{(i)}_{\ell}=\bm{x}^{(i)}_{\ell}(t)$ is the $m$-dimensional state vector for subnode $\ell$ (i.e., in layer $\ell$) of node $i$, the function $\bm{f}$ determines the dynamics of every isolated subnode, and $\bm{h}$ is the coupling function common to all sublinks.
Here, for all links of a given type between different nodes, the corresponding coupling matrix $\widetilde{A}^{(ii')} := (\widetilde{A}^{(ii')}_{\ell\ell'})$, $i\neq i'$, is the same and encodes the subnode connection pattern for that link type.
In contrast, the subnode connection pattern within each node $i$ is encoded in the matrix $\Fi := (\widetilde{A}^{(ii)}_{\ell\ell'})$.
Since the subnode-to-subnode interactions are diffusive, the synchronous state given by $\bm{x}^{(i)}_{\ell}(t)=\bm{s}(t)$, $\forall i,\ell$ with $\dot{\bm{s}} = \bm{f}\left( \bm{s} \right)$ is guaranteed to exist. 
{\TN
Note that the diffusive coupling among subnodes do not necessarily imply that the node-to-node interactions are diffusive, as intralayer synchronization of the form $\bm{x}^{(i)}_{\ell}=\bm{s}_\ell$ among subnodes is also valid as a state of complete synchronization among all nodes. The interactions among nodes do not vanish in this case due to the existence of external sublink connections among different layers.
}
To summarize, Eq.~\eqref{eq:multilayer} describes a general class of multilayer models of symmetric networks that admit a state corresponding to complete synchronization, $\bm{X}_i(t)=\bm{S}(t)$, $\forall i$, when written in the form of Eq.~\eqref{eq:0} (see Appendix B for details).

\section{Establishing \textit{AIS\MakeLowercase{ync}}}

To facilitate the stability analysis required to establish {\it AISync}, we flatten the multilayer network representation into a single layer (see Fig.~\ref{fig:1}(c) for an example).
We use $\widetilde{A} = (\widetilde{A}_{jj'})$ to denote the adjacency matrix that encodes the structure of the resulting monolayer network (see Fig.~\ref{fig:1}(d) for an example).
This matrix has a block structure in which the matrices $\Fi$ appearing on the diagonal blocks characterize node properties, while $\widetilde{A}^{(ii')}$ appearing on the off-diagonal blocks reflect the link types.
Since subnodes and sublinks are identical, we can directly apply the MSF analysis~\cite{pecora1998master} to the monolayer network and obtain the stability function $\psi(\lambda)$ (see Appendix~\ref{sec:msf} for details).  
The maximum transverse Lyapunov exponent (MTLE) of the synchronous state is then computed as $\Psi := \max_{2\leq j \leq n}\psi(\lambda_j)$, where $\lambda_1,\ldots,\lambda_n$ are the eigenvalues of 
the corresponding Laplacian matrix $\widetilde{L}:=(\widetilde{L}_{jj'})$, defined as $\widetilde{L}_{jj'} := \delta_{jj'}\sum_{k=1}^n \widetilde{A}_{jk} - \widetilde{A}_{jj'}$, where $\delta_{jj'}$ is the Kronecker delta function.
Here, $\lambda_1$ is the identically zero eigenvalue, which is excluded in the definition of $\Psi$ for corresponding to a mode of perturbation that does not affect synchronization.
Thus, the synchronous state is asymptotically stable if $\Psi < 0$ and unstable if $\Psi > 0$.

{\TN
To establish {\it AISync} for our multilayer system, we first verify that all homogeneous systems have $\Psi > 0$ (i.e., synchronous state $\bm{x}^{(i)}_{\ell}=\bm{s}$, $\forall i, \ell$, is unstable), and check numerically that all other synchronous states $\bm{x}^{(i)}_{\ell}=\bm{s}_\ell$, $\forall i, \ell$, are also unstable. This establishes condition (C1).  
We then find a heterogeneous system with $\Psi < 0$, which establishes condition (C2).
This procedure is detailed in Appendix~\ref{sec:s2}. 
}

In the case of linear $\bm{f}$ and $\bm{h}$, which is widely used to study consensus dynamics and encompasses a variety of nontrivial stability regions~\cite{li2010consensus}, the problem of verifying {\it AISync} is fully solvable.  
To see this, we first note that in this case the stability function $\psi(\lambda)$ determines the (common) stability of {\em all} completely synchronous states of the form $\bm{x}^{(i)}_{\ell}=\bm{s}_\ell$, $\forall i, \ell$, where the subnode states $\bm{s}_\ell$ can in general be different for different $\ell$.
Next, for a given (homogeneous or heterogeneous) system, we sort its Laplacian eigenvalues into two groups: $\lambda_1,\ldots,\lambda_{j^*}$, corresponding only to those perturbations parallel to the synchronization manifold, and $\lambda_{j^*+1},\ldots,\lambda_{n}$, corresponding to perturbations that are transverse to the manifold and thus destroy synchronization.
The stability (of all completely synchronous states) is then determined by $\Psi' := \max_{j^* < j \le n} \psi(\lambda_j)$, noting that both $j^*$ and $\lambda_j$ generally depend on the network structure.
This leads to the following solution for the {\it AISync} conditions: $\Psi'\geq 0$ for all homogeneous systems and $\Psi'<0$ for some heterogeneous system (where we include $\Psi'=0$ in the first condition because $\Psi' = 0$ for linear system would exclude asymptotically stable synchronization). 

\section{Examples of \textit{AIS\MakeLowercase{ync}}}

\subsection{Consensus dynamics}

{\YZ
Here we establish {\it AISync} for the system with the symmetric network structure shown in Fig.~\ref{fig:0}, in which the subnodes follow the consensus dynamics used in Ref.~\cite{li2010consensus}:
\begin{equation}
    \dot{\bm{x}}_i = D\bm{f} \, \bm{x}_i - \sum_{j} \widetilde{L}_{ij} \, D\bm{h} \, \bm{x}_j,
\end{equation}
where
\begin{equation}  
    D\bm{f} = 
     \begin{pmatrix}
      -2 & 2 & -1 & 2 \\
      -1 & 1 &  0 & 0 \\
      0  & 0 & -3 & 4 \\
      0  & 0 & -1 & 1 
     \end{pmatrix}\quad 
     D\bm{h} = 
     \begin{pmatrix}
      0 & 0 & 0 &  0 \\
      0 & 0 & 0 &  0 \\
      0 & 1 & 0 & -1 \\
     -1 & 0 & 1 &  0 
     \end{pmatrix}.
\end{equation} 
This leads to the stability region $\psi(\lambda)<0$ shown in Fig.~\ref{fig:0}(c),
defined by 
\begin{equation}
x(x+3) - y^2 - (2x+3)^2y^2 > 0,
\label{eq:stability-region-linear}
\end{equation} 
where $x$ and $y$ denote the real and imaginary parts of $\lambda$, respectively.

\begin{figure}[t]
\centering
\includegraphics[width=.9\columnwidth]{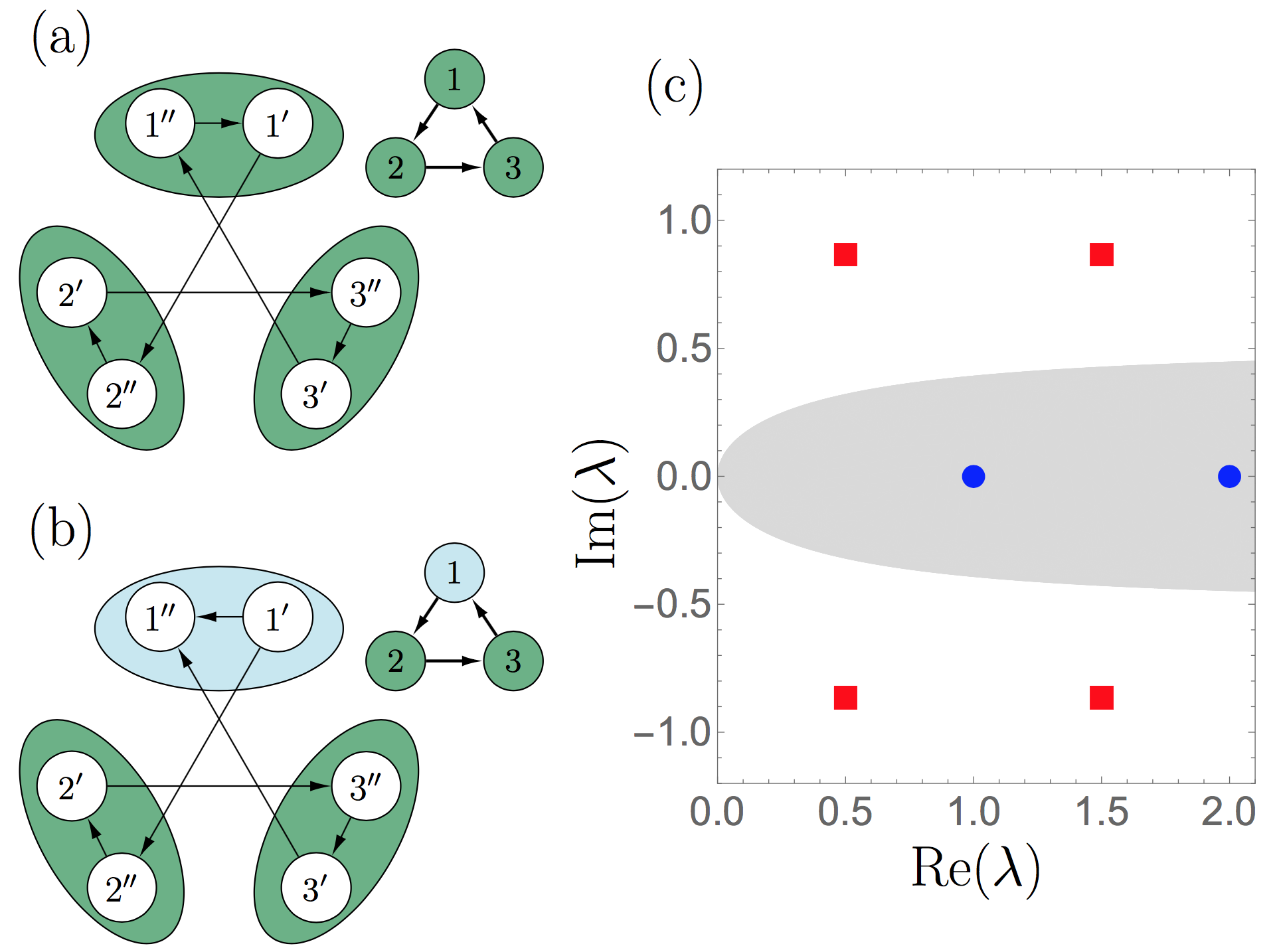}
\caption{\TN
Example of consensus system showing \textit{AISync}.
(a) Symmetric network of $N=3$ homogeneous nodes, each with $L=2$ subnodes coupled by a directed link (from subnode $i''$ to $i'$). 
(b) The same network but with heterogeneous nodes, in which the direction of the internal sublink in the (light) cyan node is the opposite of that in the (dark) green nodes. In both (a) and (b) we show the corresponding node-level visualization of the network at the top right.
(c) Stability region (shaded gray) for the consensus dynamics. All the transverse modes for the homogeneous system in (a) are unstable (red squares), while those for the heterogeneous system in (b) are stable (blue dots).
}
\label{fig:0}
\end{figure}

For $L=2$ there are only two possible homogeneous systems, associated with the two possible directions of the internal sublink in each node.
The homogeneous system in Fig.~\ref{fig:0}(a) has Laplacian eigenvalues $\lambda_1 = 0$, $\lambda_2 = 2$, $\lambda_{3,4} \approx 0.5 \pm 0.866 i$, and $\lambda_{5,6} \approx 1.5 \pm 0.866 i$, where $\lambda_1$ and $\lambda_2$ correspond to the perturbations parallel to the synchronization manifold and $\lambda_3,\ldots,\lambda_6$ correspond to those in the transverse directions (i.e., $j^* = 2$). 
Since $\psi(\lambda_j)>0$ for $j=3,4,5,6$ [i.e., all these $\lambda_j$'s fall outside the stability region defined by Eq.~\eqref{eq:stability-region-linear}, as indicated by the red squares in Fig.~\ref{fig:0}(c)],
we have $\Psi'=\max_{2<j\leq 6} \psi(\lambda_j) > 0$. 
The other homogeneous system is not synchronizable since all the single-prime subnodes have no incoming sublink.
In contrast, for the heterogeneous system in Fig.~\ref{fig:0}(b), the Laplacian eigenvalues are $\lambda_1 = 0$, $\lambda_j = 1$ for $1< j \le 5$, and $\lambda_6 = 2$ (i.e., $j^* = 1$ in this case). As shown by the blue dots in Fig.~\ref{fig:0}(c), we have $\Psi'=\max_{1<j\leq 6} \psi(\lambda_j) < 0$ for this heterogeneous system.
We thus see that $\Psi' \geq 0$ (i.e., the synchronous state is not asymptotically stable) for both homogeneous systems and $\Psi' < 0$ (i.e., the synchronous state is asymptotically stable) for a heterogeneous system, establishing {\it AISync}: the agents can reach consensus only when some of them are different from the others.
}

\subsection{Coupled Lorenz oscillators}

\begin{figure}[t]
\centering
\includegraphics[width=1\columnwidth]{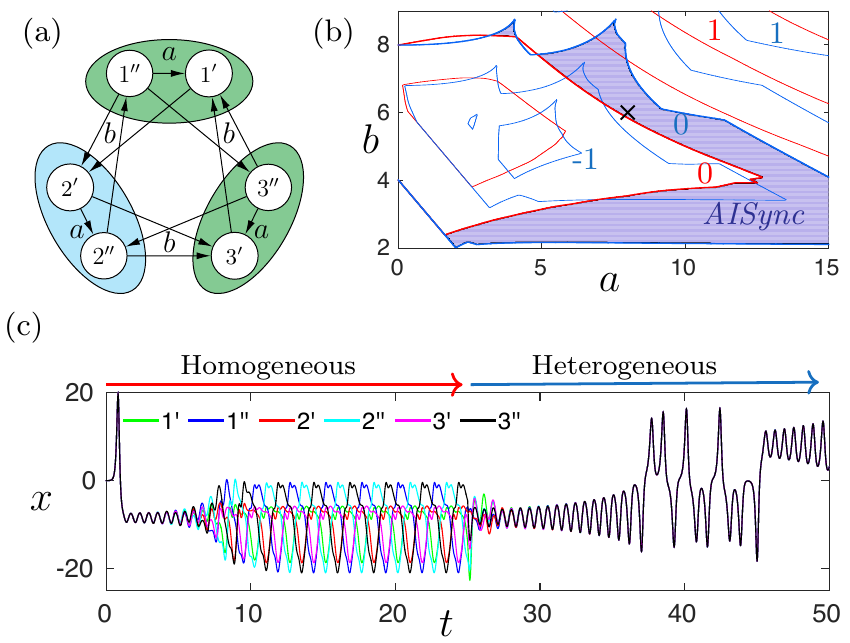}
\caption{
Example of coupled Lorenz systems showing \textit{AISync}.
(a) Symmetric network of $N=3$ nodes, each with $L=2$ directionally coupled subnodes of Lorenz oscillators.
Here we show an instance of a heterogeneous system in which the sublink direction in one node (cyan) is different from the other two (green).
(b) Contour plots of $\Psi_{=}$ (red) for the case of homogeneous nodes (all-green or all-cyan nodes) and $\Psi_{\neq}$ (blue) for heterogeneous nodes (one or two green nodes).
The shaded region corresponds to {\it AISync} systems, for which $\Psi_{=} > 0$ and $\Psi_{\neq} < 0$.
(c) Sample trajectory of the system for $a=8$ and $b=6$ [cross symbol in (b)], exhibiting {\it AISync}.
The first component $x_1$ of the Lorenz oscillator state vector is shown for all $n=6$ subnodes.
}
\label{fig:2}
\end{figure}

An example of nonlinear system exhibiting {\it AISync} is shown in Fig.~\ref{fig:2}.
The network structure is symmetric and has $N=3$ nodes and $K=2$ types of links representing sublink patterns in the clockwise and counterclockwise directions in Fig.~\ref{fig:2}(a).
Each node consists of $L=2$ subnodes, each of which is a chaotic Lorenz oscillator.
The two subnodes are connected by a sublink, the direction of which determines the node type.
This gives rise to two node types, and there are four possible distinct combinations of node types for the network---two homogeneous and two heterogeneous.
The system has two parameters, $a$ and $b$, representing the coupling strength of internal and external sublinks, respectively.
We seek to determine for which values of $a$ and $b$ the system exhibits {\it AISync}.

In Fig.~\ref{fig:2}(b), we show $\Psi_{=}$ (red) and $\Psi_{\neq}$ (blue) as functions of $a$ and $b$, where $\Psi_{=}$ ($\Psi_{\neq}$) are defined to be the smaller value of $\Psi$ between the two possible homogeneous (heterogeneous) systems.
{\TN 
In the region shaded purple (where $\Psi_{\neq} >0$ and $\Psi_{=} <0$),  the synchronous state $\bm{x}^{(i)}_{\ell}(t)=\bm{s}(t)$, $\forall i,\ell$ is stable for at least one of the heterogeneous systems, but unstable for both homogeneous systems. We further verify in this region that the other possible forms of synchronous states, $\bm{x}^{(i)}_{\ell}(t)=\bm{s}_\ell(t)$, $\forall i, \ell$, are unstable for both homogeneous systems (through extensive numerical simulation---see Appendix~\ref{sec:msf-y-x-coupling} for details). This establishes conditions (C1) and (C2), thus confirming that the system exhibits {\it AISync} in the purple region.
}
The {\it AISync} behavior of the system for a specific combination of $a$ and $b$ is illustrated by the sample trajectory in Fig.~\ref{fig:2}(c), which diverges from synchrony while the nodes are kept homogeneous, but re-synchronizes spontaneously after the nodes are made heterogeneous. 

\begin{figure*}[hbt!]
\includegraphics[width=.9\textwidth]{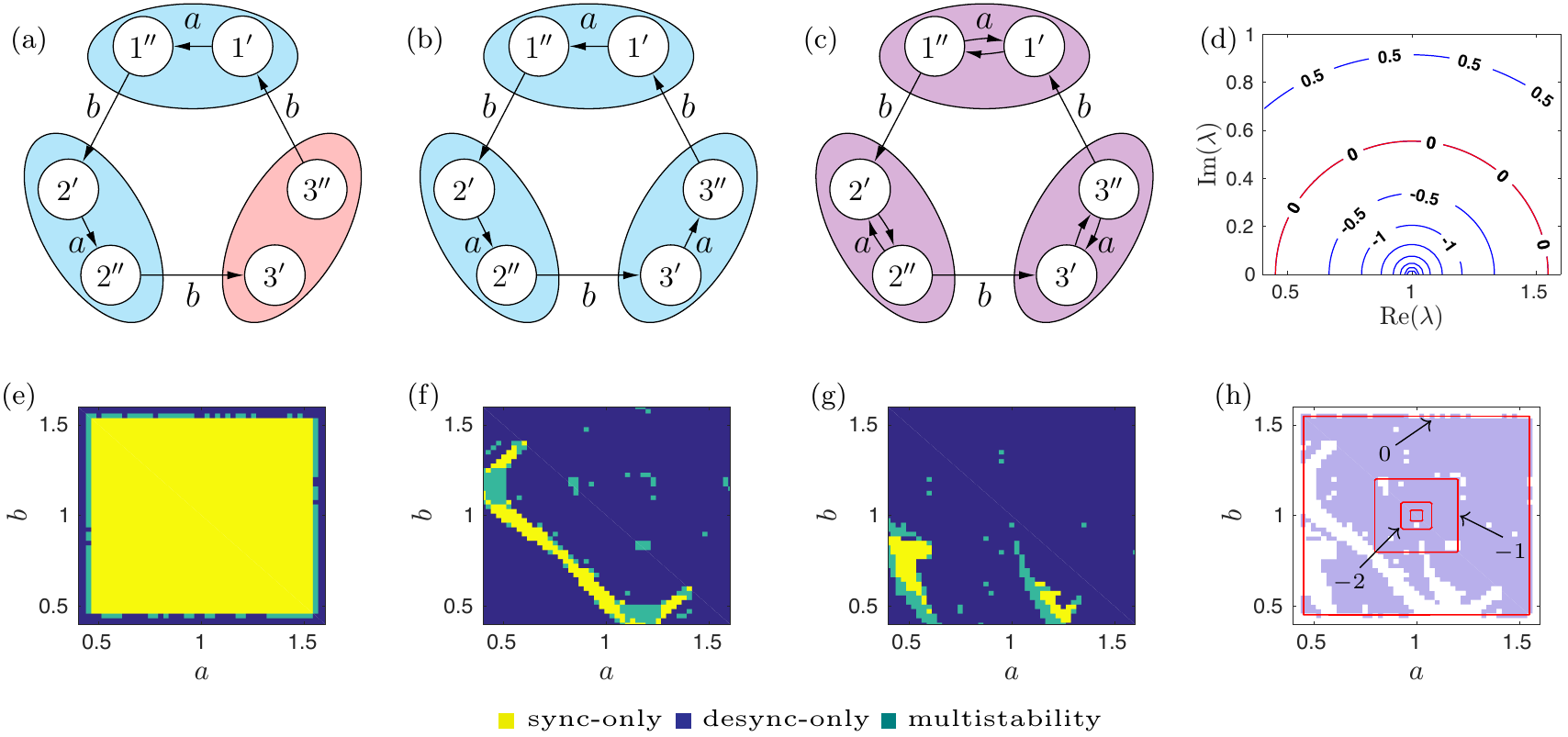}
\caption{
Example of coupled electro-optic systems showing \textit{AISync}.
(a--c) Networks with heterogeneous (a) and homogeneous nodes (b,c). 
(d) Stability function $\psi(\lambda)$ for the electro-optic system.
(e--g) Numerical results indicating where each system in (a--c) is synchronizable in the parameter space. 
Each pixel is categorized into three classes according to $24$ independent simulations from random initial conditions (see text for details).
(h) The \textit{AISync} region (shaded purple), which is the union of yellow and green regions in (e) minus the analogous unions in (f) and (g).  The red contours encode the MTLE for the heterogeneous system.
}
\label{fig:3}
\end{figure*}

\subsection{Coupled electro-optic systems}

{\YZ
We now present an experimentally testable {\it AISync} system based on the discrete-time model of the electro-optic system implemented in Refs.~\cite{hagerstrom2012experimental, pecora2014cluster} and given by
\begin{equation}
    x_i^{t+1} = \biggl[f(x_i^t) - \sum_j \widetilde{L}_{ij} f(x_j^t) + \delta \biggr] \mod 2\pi,
\label{eq:s3}
\end{equation}
where $f(x)=\beta I(x)$ determines the isolated subnode dynamics and also serves as the coupling function.
Here, $I(x) = (1 - \cos x)/2$ is the normalized optical intensity,
$\beta = 1.7\pi$ is the self-feedback strength,
$\delta = 0.2$ is the offset introduced to suppress the trivial solution $x_i=0$, and $\widetilde{L}_{ij}$ is the weighted graph Laplacian [weights controlled by parameters $a$ and $b$, as shown in Fig.~\ref{fig:3}(a--c)].

Figure~\ref{fig:3} shows an example of \textit{AISync} using these electro-optic maps as subnodes. The internal connections are chosen from the quaternary set (no sublink, one directed sublink in either direction, and directed sublinks in both directions).
When the same choice is made for all internal connections, this leads to four different homogeneous systems, but two of them have $\lambda_2=0$ (not synchronizable), leaving only two homogeneous systems to consider [Figs.~\ref{fig:3}(b) and (c)].
For comparison, we take the heterogeneous system in Fig.~\ref{fig:3}(a), 
which forms a directed chain network in its monolayer representation.
Each of the three systems [Figs.~\ref{fig:3}(a--c)] has a companion plot showing under what parameters the nodes are synchronizable 
[Figs.~\ref{fig:3}(e--g)].
In the latter panels, each pixel is generated from $24$ independent simulations run from random initial conditions. 
The pixels are then color-coded according to how many times a fully synchronized state was reached after 2500 iterations (``sync-only'': 24 times; ``desync-only'': 0 times; ``multistability'': all other cases).
Here we consider a trajectory to be fully synchronized if the synchronization error $e$ 
defined in Eq.~\eqref{eqn:error} and
averaged over the last $100$ iterations falls below $10^{-3}$.
It is worth noting that, in this example, when a homogeneous system is synchronizable the synchronous state is always in the form of cluster synchronization among subnodes (those indexed with prime and double prime
form two separate synchronized clusters), since complete synchronization among subnodes is always unstable for both homogeneous systems.

Figure~\ref{fig:3}(d) shows the stability function $\psi(\lambda)$ for the electro-optic subnode dynamics and coupling function, 
which has a bounded stable region. The lines are quite dense inside the stable region, meaning that the stability landscape is steep there and the function reaches very deep negative values. This is confirmed in Fig.~\ref{fig:3}(h), where the \textit{AISync} regions are shaded purple, with the MTLE of the synchronous state for the heterogeneous system shown as red contour lines.
}

\subsection{Other examples}

The three {\it AISync} systems considered in this section include both linear and nonlinear coupling functions, as well as discrete- and continuous-time dynamics. While they provide {\it AISync} examples with directed network structures and bounded stability regions, we also demonstrate {\it AISync} for systems with unbounded stability region (Supplemental Material Sec.~S1), and for systems with undirected network structure (Supplemental Material Sec.~S2).

\section{Propensity for \textit{AIS\MakeLowercase{ync}}}

But how often does a network structure support {\it AISync}?
To systematically address this question, we use the spread $\sigma$ of the eigenvalues of the Laplacian matrix $\widetilde{L}$ for the monolayer network representation, which is a measure of synchronizability~\cite{nishikawa2010network} defined by
$\sigma^2 := \sum_{j=2}^n |\lambda_j - \overline{\lambda}|^2 / [d^2(n-1)]$,
where $d := \sum_{j=1}^n \widetilde{L}_{jj} / n$ and $\overline{\lambda} := \sum_{j=2}^n \lambda_j / (n-1)$.
A smaller $\sigma$ indicates higher synchronizability.
Given an external sublink structure corresponding to a symmetric network,
we compare the minimum spread $\sigma_{=}$ among all systems with homogeneous $\Fi$ to the corresponding minimum $\sigma_{\neq}$ among all systems with strictly heterogeneous $\Fi$.
We call the structure {\it AISync-favoring} if $\sigma_{\neq} < \sigma_{=}$, which indicates that heterogeneous $\Fi$ can make the system easier to synchronize than any homogeneous $\Fi$.
As a measure of how strongly the structure supports {\it AISync}, we define $r := 1-\sigma_{\neq}/\sigma_{=} \le 1$, where $r>0$ indicates an {\it AISync}-favoring structure, and $r=1$ implies $\sigma_{\neq}=0$ (i.e., there is a heterogeneous system with optimal synchronizability). 
For example, the structure in Fig.~\ref{fig:2} has $\sigma_{=}\approx0.56$ and $\sigma_{\neq}\approx0.33$, and $r\approx0.41$.

\begin{table}[t]
\begin{center}
\begin{tabular}{ M{1.8cm} | M{1.6cm} | M{3cm} | M{1.6cm} }
    & $N=3$ & $N=4$ & $N=5$ \\ \hline
    symmetric \newline networks
    &
    \resizebox{.09\textwidth}{!}{
    \begin{tikzpicture}[
      vertex/.style={draw,circle,very thick,minimum size=.6cm},
      arc/.style={draw,very thick,-{Latex[length=3mm, width=2mm]},bend left=10}]
      \node[vertex] (p1) at (0,2.2) {};
      \node[vertex] (p2) at (-1.2,0) {};
      \node[vertex] (p3) at (1.2,0) {};   
      \foreach [count=\r] \row in 
      {{0,1,2},
       {2,0,1},
       {1,2,0}}
      {
          \foreach [count=\c] \cell in \row
          {
              \ifnum\cell=1
                  \draw[arc] (p\r) edge (p\c);
              \fi
              \ifnum\cell=2
                  \draw[arc,cyan] (p\r) edge (p\c);
              \fi
          }
      }
      \end{tikzpicture}
      }
      &
      \resizebox{.075\textwidth}{!}{
      \begin{tikzpicture}[
      vertex/.style={draw,circle,very thick,minimum size=.6cm},
      arc/.style={draw,very thick,-{Latex[length=3mm, width=2mm]}}]
      \node[vertex] (p1) at (1,1) {};
      \node[vertex] (p2) at (1,-1) {};
      \node[vertex] (p3) at (-1,-1) {};
      \node[vertex] (p4) at (-1,1) {};    
      \foreach [count=\r] \row in 
      {{0,-1,2,1},
       {1,0,-1,2},
       {2,1,0,-1},
       {-1,2,1,0}}
      {
          \foreach [count=\c] \cell in \row
          {
              \ifnum\cell=1
                  \draw[arc,bend left=10] (p\r) edge (p\c);
              \fi
              \ifnum\cell=-1
                  \draw[arc,red,bend left=10] (p\r) edge (p\c);
              \fi
              \ifnum\cell=2
                  \draw[arc,cyan] (p\r) edge (p\c);
              \fi
          }
      }
      \end{tikzpicture}
      }
      \resizebox{.075\textwidth}{!}{
      \begin{tikzpicture}[
      vertex/.style={draw,circle,very thick,minimum size=.6cm},
      arc/.style={draw,very thick,-{Latex[length=3mm, width=2mm]}}]
      \node[vertex] (p1) at (1,1) {};
      \node[vertex] (p2) at (1,-1) {};
      \node[vertex] (p3) at (-1,-1) {};
      \node[vertex] (p4) at (-1,1) {};    
      \foreach [count=\r] \row in 
      {{0,-1,2,1},
       {-1,0,1,2},
       {2,1,0,-1},
       {1,2,-1,0}}
      {
          \foreach [count=\c] \cell in \row
          {
              \ifnum\cell=1
                  \draw[arc] (p\r) edge (p\c);
              \fi
              \ifnum\cell=-1
                  \draw[arc,red] (p\r) edge (p\c);
              \fi
              \ifnum\cell=2
                  \draw[arc,cyan] (p\r) edge (p\c);
              \fi
          }
      }
      \end{tikzpicture}
      }
      &
      \vspace{1mm}
      \resizebox{.09\textwidth}{!}{
      \begin{tikzpicture}[
      vertex/.style={draw,circle,very thick,minimum size=.6cm},
      arc/.style={draw,very thick,-{Latex[length=3mm, width=2mm]},bend left=10}]
      \node[vertex] (p1) at (1.4,0) {};
      \node[vertex] (p2) at (.4,1.3) {};
      \node[vertex] (p3) at (-1.1,.8) {};
      \node[vertex] (p4) at (-1.1,-.8) {};
      \node[vertex] (p5) at (.4,-1.3) {}; 
      \foreach [count=\r] \row in 
      {{0,1,2,3,4},
       {4,0,1,2,3},
       {3,4,0,1,2},
       {2,3,4,0,1},
       {1,2,3,4,0}}
      {
          \foreach [count=\c] \cell in \row
          {
              \ifnum\cell=1
                  \draw[arc] (p\r) edge (p\c);
              \fi
              \ifnum\cell=2
                  \draw[arc,red] (p\r) edge (p\c);
              \fi
              \ifnum\cell=3
                  \draw[arc,cyan] (p\r) edge (p\c);
              \fi
              \ifnum\cell=4
                  \draw[arc,violet] (p\r) edge (p\c);
              \fi
          }
      }
      \end{tikzpicture}
      }\\ \hline
      Q\,(optimal)\rule{2pt}{0pt}\rule{0pt}{9pt} & 9 & 14 & 21 \\ 
      Q\,($r>0.2$)\rule{4.5pt}{0pt} & 11 & 81 & 254 \\ 
      Q\,($r>0.05$) & 29 & 318 & 2154 \\ 
      B\,($r>0.2$)\rule{4pt}{0pt}  & 11 & 101 & 204 \\ 
      B\,($r>0.05$)  & 31 & 400 & 2406 \\ 
    \end{tabular}
\vspace{-1mm}
\caption{
Number of isomorphically distinct \textit{AISync}-favoring networks, listed for $N=3,4,5$ nodes and $L=2$ layers (with $a=b=1$ to enable counting).
The numbers are given for both binary (B) and quaternary (Q) choices of internal sublink configurations, as well as for different {\it AISync} strength (as measured by $r$ defined 
in the text).
The network diagrams encode all possible symmetric networks of a given size.
}
\end{center}
\label{tbl:tbl1}
\end{table}

\begin{table*}[ht]
\begin{center}
\caption{Diagrams of symmetric networks with $N=6$, $7$, and $8$ nodes.}
{\renewcommand{\arraystretch}{1.1}
\begin{tabular}{ | M{1.6cm} | M{.22\textwidth} M{.67\textwidth} | }

    \hline
    $N=6$
    &
    \vspace{2mm}
    \resizebox{.23\textwidth}{!}{
    \begin{tikzpicture}[
      vertex/.style={draw,circle,minimum size=.3cm},
      arc/.style={draw,-{Latex[length=2mm, width=1mm]},bend left=10}]
      \node[vertex] (p1) at (.5,.86) {};
      \node[vertex] (p2) at (-.5,.86) {};
      \node[vertex] (p3) at (-1,0) {}; 
      \node[vertex] (p4) at (-.5,-.86) {};
      \node[vertex] (p5) at (.5,-.86) {};
      \node[vertex] (p6) at (1,0) {}; 
      \node[] (l1) at (1.5,0) {$=$}; 
      \foreach [count=\r] \row in 
      {{0,1,3,5,4,2},
       {2,0,1,3,5,4},
       {4,2,0,1,3,5},
       {5,4,2,0,1,3},
       {3,5,4,2,0,1},
       {1,3,5,4,2,0}}
      {
          \foreach [count=\c] \cell in \row
          {
              \ifnum\cell=1
                  \draw[arc] (p\r) edge (p\c);
              \fi
              \ifnum\cell=2
                  \draw[arc,red] (p\r) edge (p\c);
              \fi
              \ifnum\cell=3
                  \draw[arc,cyan] (p\r) edge (p\c);
              \fi
              \ifnum\cell=4
                  \draw[arc,violet] (p\r) edge (p\c);
              \fi
              \ifnum\cell=5
                  \draw[arc,ForestGreen,bend left=0] (p\r) edge (p\c);
              \fi
          }
      }
      \end{tikzpicture}
      }
    \resizebox{.23\textwidth}{!}{
    \begin{tikzpicture}[
      vertex/.style={draw,circle,minimum size=.3cm},
      arc/.style={draw,-{Latex[length=2mm, width=1mm]},bend left=10}]
      \node[vertex] (p1) at (.5,.86) {};
      \node[vertex] (p2) at (-.5,.86) {};
      \node[vertex] (p3) at (-1,0) {}; 
      \node[vertex] (p4) at (-.5,-.86) {};
      \node[vertex] (p5) at (.5,-.86) {};
      \node[vertex] (p6) at (1,0) {};
      \node[] (l1) at (1.5,0) {$=$};
      \foreach [count=\r] \row in 
      {{0,1,3,5,4,2},
       {1,0,2,3,5,4},
       {4,2,0,1,3,5},
       {5,4,1,0,2,3},
       {3,5,4,2,0,1},
       {2,3,5,4,1,0}}
      {
          \foreach [count=\c] \cell in \row
          {
              \ifnum\cell=1
                  \draw[arc,bend left=0] (p\r) edge (p\c);
              \fi
              \ifnum\cell=2
                  \draw[arc,red,bend left=0] (p\r) edge (p\c);
              \fi
              \ifnum\cell=3
                  \draw[arc,cyan] (p\r) edge (p\c);
              \fi
              \ifnum\cell=4
                  \draw[arc,violet] (p\r) edge (p\c);
              \fi
              \ifnum\cell=5
                  \draw[arc,ForestGreen,bend left=0] (p\r) edge (p\c);
              \fi
          }
      }
      \end{tikzpicture}
      }
    \resizebox{.23\textwidth}{!}{
    \begin{tikzpicture}[
      vertex/.style={draw,circle,minimum size=.3cm},
      arc/.style={draw,-{Latex[length=2mm, width=1mm]},bend left=10}]
      \node[vertex] (p1) at (.5,.86) {};
      \node[vertex] (p2) at (-.5,.86) {};
      \node[vertex] (p3) at (-1,0) {}; 
      \node[vertex] (p4) at (-.5,-.86) {};
      \node[vertex] (p5) at (.5,-.86) {};
      \node[vertex] (p6) at (1,0) {}; 
      \node[] (l1) at (1.5,0) {$=$}; 
      \foreach [count=\r] \row in 
      {{0,1,4,5,3,2},
       {2,0,1,3,5,4},
       {3,2,0,1,4,5},
       {5,4,2,0,1,3},
       {4,5,3,2,0,1},
       {1,3,5,4,2,0}}
      {
          \foreach [count=\c] \cell in \row
          {
              \ifnum\cell=1
                  \draw[arc] (p\r) edge (p\c);
              \fi
              \ifnum\cell=2
                  \draw[arc,red] (p\r) edge (p\c);
              \fi
              \ifnum\cell=3
                  \draw[arc,cyan] (p\r) edge (p\c);
              \fi
              \ifnum\cell=4
                  \draw[arc,violet] (p\r) edge (p\c);
              \fi
              \ifnum\cell=5
                  \draw[arc,ForestGreen,bend left=0] (p\r) edge (p\c);
              \fi
          }
      }
      \end{tikzpicture}
      }
    \resizebox{.23\textwidth}{!}{
    \begin{tikzpicture}[
      vertex/.style={draw,circle,minimum size=.3cm},
      arc/.style={draw,-{Latex[length=2mm, width=1mm]},bend left=10}]
      \node[vertex] (p1) at (.5,.86) {};
      \node[vertex] (p2) at (-.5,.86) {};
      \node[vertex] (p3) at (-1,0) {}; 
      \node[vertex] (p4) at (-.5,-.86) {};
      \node[vertex] (p5) at (.5,-.86) {};
      \node[vertex] (p6) at (1,0) {}; 
      \node[] (l1) at (1.5,0) {$=$}; 
      \foreach [count=\r] \row in 
      {{0,1,4,5,3,2},
       {1,0,2,3,5,4},
       {3,2,0,1,4,5},
       {5,4,1,0,2,3},
       {4,5,3,2,0,1},
       {2,3,5,4,1,0}}
      {
          \foreach [count=\c] \cell in \row
          {
              \ifnum\cell=1
                  \draw[arc,bend left=0] (p\r) edge (p\c);
              \fi
              \ifnum\cell=2
                  \draw[arc,red,bend left=0] (p\r) edge (p\c);
              \fi
              \ifnum\cell=3
                  \draw[arc,cyan] (p\r) edge (p\c);
              \fi
              \ifnum\cell=4
                  \draw[arc,violet] (p\r) edge (p\c);
              \fi
              \ifnum\cell=5
                  \draw[arc,ForestGreen,bend left=0] (p\r) edge (p\c);
              \fi
          }
      }
      \end{tikzpicture}
      }

      &

    \vspace{-1mm}
    \resizebox{.23\textwidth}{!}{
    \begin{tikzpicture}[
      vertex/.style={draw,circle,minimum size=.3cm},
      arc/.style={draw,-{Latex[length=2mm, width=1mm]},bend left=10}]
      \draw [fill=SkyBlue!15,SkyBlue!15] (-1.1,-1) rectangle (1.1,1);
      \node[vertex] (p1) at (.5,.86) {};
      \node[vertex] (p2) at (-.5,.86) {};
      \node[vertex] (p3) at (-1,0) {}; 
      \node[vertex] (p4) at (-.5,-.86) {};
      \node[vertex] (p5) at (.5,-.86) {};
      \node[vertex] (p6) at (1,0) {}; 
      \node[] (l1) at (1.5,0) {$+$}; 
      \foreach [count=\r] \row in 
      {{0,1,3,5,4,2},
       {2,0,1,3,5,4},
       {4,2,0,1,3,5},
       {5,4,2,0,1,3},
       {3,5,4,2,0,1},
       {1,3,5,4,2,0}}
      {
          \foreach [count=\c] \cell in \row
          {
              \ifnum\cell=3
                  \draw[arc,Gray!40] (p\r) edge (p\c);
              \fi
              \ifnum\cell=4
                  \draw[arc,Gray!40] (p\r) edge (p\c);
              \fi
          }
      }
      \foreach [count=\r] \row in 
      {{0,1,3,5,4,2},
       {2,0,1,3,5,4},
       {4,2,0,1,3,5},
       {5,4,2,0,1,3},
       {3,5,4,2,0,1},
       {1,3,5,4,2,0}}
      {
          \foreach [count=\c] \cell in \row
          {
              \ifnum\cell=1
                  \draw[arc] (p\r) edge (p\c);
              \fi
              \ifnum\cell=2
                  \draw[arc,red] (p\r) edge (p\c);
              \fi
              \ifnum\cell=5
                  \draw[arc,ForestGreen,bend left=0] (p\r) edge (p\c);
              \fi
          }
      }
      \end{tikzpicture}
      }
    \resizebox{.23\textwidth}{!}{
    \begin{tikzpicture}[
      vertex/.style={draw,circle,minimum size=.3cm},
      arc/.style={draw,-{Latex[length=2mm, width=1mm]},bend left=10}]
      \node[vertex] (p1) at (.5,.86) {};
      \node[vertex] (p2) at (-.5,.86) {};
      \node[vertex] (p3) at (-1,0) {}; 
      \node[vertex] (p4) at (-.5,-.86) {};
      \node[vertex] (p5) at (.5,-.86) {};
      \node[vertex] (p6) at (1,0) {};  
      \node[] (l1) at (1.5,0) {$+$};
      \foreach [count=\r] \row in 
      {{0,1,3,5,4,2},
       {2,0,1,3,5,4},
       {4,2,0,1,3,5},
       {5,4,2,0,1,3},
       {3,5,4,2,0,1},
       {1,3,5,4,2,0}}
      {
          \foreach [count=\c] \cell in \row
          {
              \ifnum\cell=1
                  \draw[arc,Gray!40] (p\r) edge (p\c);
              \fi
              \ifnum\cell=2
                  \draw[arc,Gray!40] (p\r) edge (p\c);
              \fi
              \ifnum\cell=4
                  \draw[arc,Gray!40] (p\r) edge (p\c);
              \fi
              \ifnum\cell=5
                  \draw[arc,Gray!40,bend left=0] (p\r) edge (p\c);
              \fi
          }
      }
      \foreach [count=\r] \row in 
      {{0,1,3,5,4,2},
       {2,0,1,3,5,4},
       {4,2,0,1,3,5},
       {5,4,2,0,1,3},
       {3,5,4,2,0,1},
       {1,3,5,4,2,0}}
      {
          \foreach [count=\c] \cell in \row
          {
              \ifnum\cell=3
                  \draw[arc,cyan] (p\r) edge (p\c);
              \fi
          }
      }
      \end{tikzpicture}
      }
    \resizebox{.188\textwidth}{!}{
    \begin{tikzpicture}[
      vertex/.style={draw,circle,minimum size=.3cm},
      arc/.style={draw,-{Latex[length=2mm, width=1mm]},bend left=10}]
      \node[vertex] (p1) at (.5,.86) {};
      \node[vertex] (p2) at (-.5,.86) {};
      \node[vertex] (p3) at (-1,0) {}; 
      \node[vertex] (p4) at (-.5,-.86) {};
      \node[vertex] (p5) at (.5,-.86) {};
      \node[vertex] (p6) at (1,0) {};  
      \foreach [count=\r] \row in 
      {{0,1,3,5,4,2},
       {1,0,2,3,5,4},
       {4,2,0,1,3,5},
       {5,4,1,0,2,3},
       {3,5,4,2,0,1},
       {2,3,5,4,1,0}}
      {
          \foreach [count=\c] \cell in \row
          {
              \ifnum\cell=1
                  \draw[arc,Gray!40] (p\r) edge (p\c);
              \fi
              \ifnum\cell=2
                  \draw[arc,Gray!40] (p\r) edge (p\c);
              \fi
              \ifnum\cell=3
                  \draw[arc,Gray!40] (p\r) edge (p\c);
              \fi
              \ifnum\cell=5
                  \draw[arc,Gray!40,bend left=0] (p\r) edge (p\c);
              \fi
          }
      }
      \foreach [count=\r] \row in 
      {{0,1,3,5,4,2},
       {1,0,2,3,5,4},
       {4,2,0,1,3,5},
       {5,4,1,0,2,3},
       {3,5,4,2,0,1},
       {2,3,5,4,1,0}}
      {
          \foreach [count=\c] \cell in \row
          {
              \ifnum\cell=4
                  \draw[arc,violet] (p\r) edge (p\c);
              \fi
          }
      }
      \end{tikzpicture}
      }
    
    \resizebox{.23\textwidth}{!}{
    \begin{tikzpicture}[
      vertex/.style={draw,circle,minimum size=.3cm},
      arc/.style={draw,-{Latex[length=2mm, width=1mm]},bend left=10}]
      \node[vertex] (p1) at (.5,.86) {};
      \node[vertex] (p2) at (-.5,.86) {};
      \node[vertex] (p3) at (-1,0) {}; 
      \node[vertex] (p4) at (-.5,-.86) {};
      \node[vertex] (p5) at (.5,-.86) {};
      \node[vertex] (p6) at (1,0) {}; 
      \node[] (l1) at (1.5,0) {$+$}; 
      \foreach [count=\r] \row in 
      {{0,1,3,5,4,2},
       {1,0,2,3,5,4},
       {4,2,0,1,3,5},
       {5,4,1,0,2,3},
       {3,5,4,2,0,1},
       {2,3,5,4,1,0}}
      {
          \foreach [count=\c] \cell in \row
          {
              \ifnum\cell=3
                  \draw[arc,Gray!40] (p\r) edge (p\c);
              \fi
              \ifnum\cell=4
                  \draw[arc,Gray!40] (p\r) edge (p\c);
              \fi
          }
      }
      \foreach [count=\r] \row in 
      {{0,1,3,5,4,2},
       {1,0,2,3,5,4},
       {4,2,0,1,3,5},
       {5,4,1,0,2,3},
       {3,5,4,2,0,1},
       {2,3,5,4,1,0}}
      {
          \foreach [count=\c] \cell in \row
          {
              \ifnum\cell=1
                  \draw[arc,bend left=0] (p\r) edge (p\c);
              \fi
              \ifnum\cell=2
                  \draw[arc,red,bend left=0] (p\r) edge (p\c);
              \fi
              \ifnum\cell=5
                  \draw[arc,ForestGreen,bend left=0] (p\r) edge (p\c);
              \fi
          }
      }
      \end{tikzpicture}
      }
    \resizebox{.23\textwidth}{!}{
    \begin{tikzpicture}[
      vertex/.style={draw,circle,minimum size=.3cm},
      arc/.style={draw,-{Latex[length=2mm, width=1mm]},bend left=10}]
      \node[vertex] (p1) at (.5,.86) {};
      \node[vertex] (p2) at (-.5,.86) {};
      \node[vertex] (p3) at (-1,0) {}; 
      \node[vertex] (p4) at (-.5,-.86) {};
      \node[vertex] (p5) at (.5,-.86) {};
      \node[vertex] (p6) at (1,0) {}; 
      \node[] (l1) at (1.5,0) {$+$}; 
      \foreach [count=\r] \row in 
      {{0,1,3,5,4,2},
       {1,0,2,3,5,4},
       {4,2,0,1,3,5},
       {5,4,1,0,2,3},
       {3,5,4,2,0,1},
       {2,3,5,4,1,0}}
      {
          \foreach [count=\c] \cell in \row
          {
              \ifnum\cell=1
                  \draw[arc,Gray!40,bend left=0] (p\r) edge (p\c);
              \fi
              \ifnum\cell=2
                  \draw[arc,Gray!40,bend left=0] (p\r) edge (p\c);
              \fi
              \ifnum\cell=4
                  \draw[arc,Gray!40] (p\r) edge (p\c);
              \fi
              \ifnum\cell=5
                  \draw[arc,Gray!40,bend left=0] (p\r) edge (p\c);
              \fi
          }
      }
      \foreach [count=\r] \row in 
      {{0,1,3,5,4,2},
       {1,0,2,3,5,4},
       {4,2,0,1,3,5},
       {5,4,1,0,2,3},
       {3,5,4,2,0,1},
       {2,3,5,4,1,0}}
      {
          \foreach [count=\c] \cell in \row
          {
              \ifnum\cell=3
                  \draw[arc,cyan] (p\r) edge (p\c);
              \fi
          }
      }
      \end{tikzpicture}
      }
    \resizebox{.188\textwidth}{!}{
    \begin{tikzpicture}[
      vertex/.style={draw,circle,minimum size=.3cm},
      arc/.style={draw,-{Latex[length=2mm, width=1mm]},bend left=10}]
      \node[vertex] (p1) at (.5,.86) {};
      \node[vertex] (p2) at (-.5,.86) {};
      \node[vertex] (p3) at (-1,0) {}; 
      \node[vertex] (p4) at (-.5,-.86) {};
      \node[vertex] (p5) at (.5,-.86) {};
      \node[vertex] (p6) at (1,0) {};  
      \foreach [count=\r] \row in 
      {{0,1,3,5,4,2},
       {1,0,2,3,5,4},
       {4,2,0,1,3,5},
       {5,4,1,0,2,3},
       {3,5,4,2,0,1},
       {2,3,5,4,1,0}}
      {
          \foreach [count=\c] \cell in \row
          {
              \ifnum\cell=1
                  \draw[arc,Gray!40] (p\r) edge (p\c);
              \fi
              \ifnum\cell=2
                  \draw[arc,Gray!40] (p\r) edge (p\c);
              \fi
              \ifnum\cell=3
                  \draw[arc,Gray!40] (p\r) edge (p\c);
              \fi
              \ifnum\cell=5
                  \draw[arc,Gray!40,bend left=0] (p\r) edge (p\c);
              \fi
          }
      }
      \foreach [count=\r] \row in 
      {{0,1,3,5,4,2},
       {1,0,2,3,5,4},
       {4,2,0,1,3,5},
       {5,4,1,0,2,3},
       {3,5,4,2,0,1},
       {2,3,5,4,1,0}}
      {
          \foreach [count=\c] \cell in \row
          {
              \ifnum\cell=4
                  \draw[arc,violet] (p\r) edge (p\c);
              \fi
          }
      }
      \end{tikzpicture}
      }

    \resizebox{.23\textwidth}{!}{
    \begin{tikzpicture}[
      vertex/.style={draw,circle,minimum size=.3cm},
      arc/.style={draw,-{Latex[length=2mm, width=1mm]},bend left=10}]
      \draw [fill=SkyBlue!15,SkyBlue!15] (-1.1,-1) rectangle (1.1,1);
      \node[vertex] (p1) at (.5,.86) {};
      \node[vertex] (p2) at (-.5,.86) {};
      \node[vertex] (p3) at (-1,0) {}; 
      \node[vertex] (p4) at (-.5,-.86) {};
      \node[vertex] (p5) at (.5,-.86) {};
      \node[vertex] (p6) at (1,0) {}; 
      \node[] (l1) at (1.5,0) {$+$}; 
      \foreach [count=\r] \row in 
      {{0,1,4,5,3,2},
       {2,0,1,3,5,4},
       {3,2,0,1,4,5},
       {5,4,2,0,1,3},
       {4,5,3,2,0,1},
       {1,3,5,4,2,0}}
      {
          \foreach [count=\c] \cell in \row
          {
              \ifnum\cell=3
                  \draw[arc,Gray!40] (p\r) edge (p\c);
              \fi
              \ifnum\cell=4
                  \draw[arc,Gray!40] (p\r) edge (p\c);
              \fi
          }
      }
      \foreach [count=\r] \row in
      {{0,1,4,5,3,2},
       {2,0,1,3,5,4},
       {3,2,0,1,4,5},
       {5,4,2,0,1,3},
       {4,5,3,2,0,1},
       {1,3,5,4,2,0}}
      {
          \foreach [count=\c] \cell in \row
          {
              \ifnum\cell=1
                  \draw[arc] (p\r) edge (p\c);
              \fi
              \ifnum\cell=2
                  \draw[arc,red] (p\r) edge (p\c);
              \fi
              \ifnum\cell=5
                  \draw[arc,ForestGreen,bend left=0] (p\r) edge (p\c);
              \fi
          }
      }
      \end{tikzpicture}
      }
    \resizebox{.23\textwidth}{!}{
    \begin{tikzpicture}[
      vertex/.style={draw,circle,minimum size=.3cm},
      arc/.style={draw,-{Latex[length=2mm, width=1mm]},bend left=10}]
      \draw [fill=SkyBlue!15,SkyBlue!15] (-1.1,-1) rectangle (1.1,1);
      \node[vertex] (p1) at (.5,.86) {};
      \node[vertex] (p2) at (-.5,.86) {};
      \node[vertex] (p3) at (-1,0) {}; 
      \node[vertex] (p4) at (-.5,-.86) {};
      \node[vertex] (p5) at (.5,-.86) {};
      \node[vertex] (p6) at (1,0) {};  
      \node[] (l1) at (1.5,0) {$+$};
      \foreach [count=\r] \row in 
      {{0,1,4,5,3,2},
       {1,0,2,3,5,4},
       {3,2,0,1,4,5},
       {5,4,1,0,2,3},
       {4,5,3,2,0,1},
       {2,3,5,4,1,0}}
      {
          \foreach [count=\c] \cell in \row
          {
              \ifnum\cell=1
                  \draw[arc,Gray!40,bend left=0] (p\r) edge (p\c);
              \fi
              \ifnum\cell=2
                  \draw[arc,Gray!40,bend left=0] (p\r) edge (p\c);
              \fi
              \ifnum\cell=4
                  \draw[arc,Gray!40] (p\r) edge (p\c);
              \fi
              \ifnum\cell=5
                  \draw[arc,Gray!40,bend left=0] (p\r) edge (p\c);
              \fi
          }
      }
      \foreach [count=\r] \row in 
      {{0,1,4,5,3,2},
       {1,0,2,3,5,4},
       {3,2,0,1,4,5},
       {5,4,1,0,2,3},
       {4,5,3,2,0,1},
       {2,3,5,4,1,0}}
      {
          \foreach [count=\c] \cell in \row
          {
              \ifnum\cell=3
                  \draw[arc,cyan] (p\r) edge (p\c);
              \fi
          }
      }
      \end{tikzpicture}
      }
    \resizebox{.188\textwidth}{!}{
    \begin{tikzpicture}[
      vertex/.style={draw,circle,minimum size=.3cm},
      arc/.style={draw,-{Latex[length=2mm, width=1mm]},bend left=10}]
      \draw [fill=SkyBlue!15,SkyBlue!15] (-1.1,-1) rectangle (1.1,1);
      \node[vertex] (p1) at (.5,.86) {};
      \node[vertex] (p2) at (-.5,.86) {};
      \node[vertex] (p3) at (-1,0) {}; 
      \node[vertex] (p4) at (-.5,-.86) {};
      \node[vertex] (p5) at (.5,-.86) {};
      \node[vertex] (p6) at (1,0) {};  
      \foreach [count=\r] \row in 
      {{0,1,4,5,3,2},
       {2,0,1,3,5,4},
       {3,2,0,1,4,5},
       {5,4,2,0,1,3},
       {4,5,3,2,0,1},
       {1,3,5,4,2,0}}
      {
          \foreach [count=\c] \cell in \row
          {
              \ifnum\cell=1
                  \draw[arc,Gray!40] (p\r) edge (p\c);
              \fi
              \ifnum\cell=2
                  \draw[arc,Gray!40] (p\r) edge (p\c);
              \fi
              \ifnum\cell=3
                  \draw[arc,Gray!40] (p\r) edge (p\c);
              \fi
              \ifnum\cell=5
                  \draw[arc,Gray!40,bend left=0] (p\r) edge (p\c);
              \fi
          }
      }
      \foreach [count=\r] \row in 
      {{0,1,4,5,3,2},
       {2,0,1,3,5,4},
       {3,2,0,1,4,5},
       {5,4,2,0,1,3},
       {4,5,3,2,0,1},
       {1,3,5,4,2,0}}
      {
          \foreach [count=\c] \cell in \row
          {
              \ifnum\cell=4
                  \draw[arc,violet] (p\r) edge (p\c);
              \fi
          }
      }
      \end{tikzpicture}
      }

    \resizebox{.23\textwidth}{!}{
    \begin{tikzpicture}[
      vertex/.style={draw,circle,minimum size=.3cm},
      arc/.style={draw,-{Latex[length=2mm, width=1mm]},bend left=10}]
      \node[vertex] (p1) at (.5,.86) {};
      \node[vertex] (p2) at (-.5,.86) {};
      \node[vertex] (p3) at (-1,0) {}; 
      \node[vertex] (p4) at (-.5,-.86) {};
      \node[vertex] (p5) at (.5,-.86) {};
      \node[vertex] (p6) at (1,0) {}; 
      \node[] (l1) at (1.5,0) {$+$}; 
      \foreach [count=\r] \row in 
      {{0,1,4,5,3,2},
       {1,0,2,3,5,4},
       {3,2,0,1,4,5},
       {5,4,1,0,2,3},
       {4,5,3,2,0,1},
       {2,3,5,4,1,0}}
      {
          \foreach [count=\c] \cell in \row
          {
              \ifnum\cell=3
                  \draw[arc,Gray!40] (p\r) edge (p\c);
              \fi
              \ifnum\cell=4
                  \draw[arc,Gray!40] (p\r) edge (p\c);
              \fi
          }
      }
      \foreach [count=\r] \row in 
      {{0,1,4,5,3,2},
       {1,0,2,3,5,4},
       {3,2,0,1,4,5},
       {5,4,1,0,2,3},
       {4,5,3,2,0,1},
       {2,3,5,4,1,0}}
      {
          \foreach [count=\c] \cell in \row
          {
              \ifnum\cell=1
                  \draw[arc,bend left=0] (p\r) edge (p\c);
              \fi
              \ifnum\cell=2
                  \draw[arc,red,bend left=0] (p\r) edge (p\c);
              \fi
              \ifnum\cell=5
                  \draw[arc,ForestGreen,bend left=0] (p\r) edge (p\c);
              \fi
          }
      }
      \end{tikzpicture}
      }
    \resizebox{.23\textwidth}{!}{
    \begin{tikzpicture}[
      vertex/.style={draw,circle,minimum size=.3cm},
      arc/.style={draw,-{Latex[length=2mm, width=1mm]},bend left=10}]
      \draw [fill=SkyBlue!15,SkyBlue!15] (-1.1,-1) rectangle (1.1,1);
      \node[vertex] (p1) at (.5,.86) {};
      \node[vertex] (p2) at (-.5,.86) {};
      \node[vertex] (p3) at (-1,0) {}; 
      \node[vertex] (p4) at (-.5,-.86) {};
      \node[vertex] (p5) at (.5,-.86) {};
      \node[vertex] (p6) at (1,0) {};
      \node[] (l1) at (1.5,0) {$+$};  
      \foreach [count=\r] \row in 
      {{0,1,4,5,3,2},
       {1,0,2,3,5,4},
       {3,2,0,1,4,5},
       {5,4,1,0,2,3},
       {4,5,3,2,0,1},
       {2,3,5,4,1,0}}
      {
          \foreach [count=\c] \cell in \row
          {
              \ifnum\cell=1
                  \draw[arc,Gray!40,bend left=0] (p\r) edge (p\c);
              \fi
              \ifnum\cell=2
                  \draw[arc,Gray!40,bend left=0] (p\r) edge (p\c);
              \fi
              \ifnum\cell=4
                  \draw[arc,Gray!40] (p\r) edge (p\c);
              \fi
              \ifnum\cell=5
                  \draw[arc,Gray!40,bend left=0] (p\r) edge (p\c);
              \fi
          }
      }
      \foreach [count=\r] \row in 
      {{0,1,4,5,3,2},
       {1,0,2,3,5,4},
       {3,2,0,1,4,5},
       {5,4,1,0,2,3},
       {4,5,3,2,0,1},
       {2,3,5,4,1,0}}
      {
          \foreach [count=\c] \cell in \row
          {
              \ifnum\cell=3
                  \draw[arc,cyan] (p\r) edge (p\c);
              \fi
          }
      }
      \end{tikzpicture}
      }
    \resizebox{.188\textwidth}{!}{
    \begin{tikzpicture}[
      vertex/.style={draw,circle,minimum size=.3cm},
      arc/.style={draw,-{Latex[length=2mm, width=1mm]},bend left=10}]
      \draw [fill=SkyBlue!15,SkyBlue!15] (-1.1,-1) rectangle (1.1,1);
      \node[vertex] (p1) at (.5,.86) {};
      \node[vertex] (p2) at (-.5,.86) {};
      \node[vertex] (p3) at (-1,0) {}; 
      \node[vertex] (p4) at (-.5,-.86) {};
      \node[vertex] (p5) at (.5,-.86) {};
      \node[vertex] (p6) at (1,0) {};  
      \foreach [count=\r] \row in 
      {{0,1,4,5,3,2},
       {1,0,2,3,5,4},
       {3,2,0,1,4,5},
       {5,4,1,0,2,3},
       {4,5,3,2,0,1},
       {2,3,5,4,1,0}}
      {
          \foreach [count=\c] \cell in \row
          {
              \ifnum\cell=1
                  \draw[arc,Gray!40,bend left=0] (p\r) edge (p\c);
              \fi
              \ifnum\cell=2
                  \draw[arc,Gray!40,bend left=0] (p\r) edge (p\c);
              \fi
              \ifnum\cell=3
                  \draw[arc,Gray!40] (p\r) edge (p\c);
              \fi
              \ifnum\cell=5
                  \draw[arc,Gray!40,bend left=0] (p\r) edge (p\c);
              \fi
          }
      }
      \foreach [count=\r] \row in 
      {{0,1,4,5,3,2},
       {1,0,2,3,5,4},
       {3,2,0,1,4,5},
       {5,4,1,0,2,3},
       {4,5,3,2,0,1},
       {2,3,5,4,1,0}}
      {
          \foreach [count=\c] \cell in \row
          {
              \ifnum\cell=4
                  \draw[arc,violet] (p\r) edge (p\c);
              \fi
          }
      }
      \end{tikzpicture}
      } \\ \hline
      $N=7$ &
      \vspace{1mm}
      \resizebox{.23\textwidth}{!}{
      \begin{tikzpicture}[
      vertex/.style={draw,circle,minimum size=.3cm},
      arc/.style={draw,bend left=10,-{Latex[length=2mm, width=1mm]}}]
      \node[vertex] (p1) at (.62,.78) {};
      \node[vertex] (p2) at (-.22,.97) {};
      \node[vertex] (p3) at (-.9,.43) {};
      \node[vertex] (p4) at (-.9,-.43) {};
      \node[vertex] (p5) at (-.22,-.97) {}; 
      \node[vertex] (p6) at (.62,-.78) {};
      \node[vertex] (p7) at (1,0) {}; 
      \node[] (l1) at (1.5,0) {$=$}; 
      \foreach [count=\r] \row in 
      {{0,1,2,3,4,5,6},
       {6,0,1,2,3,4,5},
       {5,6,0,1,2,3,4},
       {4,5,6,0,1,2,3},
       {3,4,5,6,0,1,2},
       {2,3,4,5,6,0,1},
       {1,2,3,4,5,6,0}}
      {
          \foreach [count=\c] \cell in \row
          {
              \ifnum\cell=1
                  \draw[arc] (p\r) edge (p\c);
              \fi
              \ifnum\cell=2
                  \draw[arc,cyan] (p\r) edge (p\c);
              \fi
              \ifnum\cell=3
                  \draw[arc,orange] (p\r) edge (p\c);
              \fi
              \ifnum\cell=4
                  \draw[arc,ForestGreen] (p\r) edge (p\c);
              \fi
              \ifnum\cell=5
                  \draw[arc,violet] (p\r) edge (p\c);
              \fi
              \ifnum\cell=6
                  \draw[arc,red] (p\r) edge (p\c);
              \fi
          }
      }
      \end{tikzpicture}
      }
      &
      \vspace{1mm}
      \resizebox{.23\textwidth}{!}{
      \begin{tikzpicture}[
      vertex/.style={draw,circle,minimum size=.3cm},
      arc/.style={draw,bend left=10,-{Latex[length=2mm, width=1mm]}}]
      \node[vertex] (p1) at (.62,.78) {};
      \node[vertex] (p2) at (-.22,.97) {};
      \node[vertex] (p3) at (-.9,.43) {};
      \node[vertex] (p4) at (-.9,-.43) {};
      \node[vertex] (p5) at (-.22,-.97) {}; 
      \node[vertex] (p6) at (.62,-.78) {};
      \node[vertex] (p7) at (1,0) {}; 
      \node[] (l1) at (1.5,0) {$+$}; 
      \foreach [count=\r] \row in 
      {{0,1,2,3,4,5,6},
       {6,0,1,2,3,4,5},
       {5,6,0,1,2,3,4},
       {4,5,6,0,1,2,3},
       {3,4,5,6,0,1,2},
       {2,3,4,5,6,0,1},
       {1,2,3,4,5,6,0}}
      {
          \foreach [count=\c] \cell in \row
          {
              \ifnum\cell=2
                  \draw[arc,Gray!40] (p\r) edge (p\c);
              \fi
              \ifnum\cell=3
                  \draw[arc,Gray!40] (p\r) edge (p\c);
              \fi
              \ifnum\cell=4
                  \draw[arc,Gray!40] (p\r) edge (p\c);
              \fi
              \ifnum\cell=5
                  \draw[arc,Gray!40] (p\r) edge (p\c);
              \fi
          }
      }
      \foreach [count=\r] \row in 
      {{0,1,2,3,4,5,6},
       {6,0,1,2,3,4,5},
       {5,6,0,1,2,3,4},
       {4,5,6,0,1,2,3},
       {3,4,5,6,0,1,2},
       {2,3,4,5,6,0,1},
       {1,2,3,4,5,6,0}}
      {
          \foreach [count=\c] \cell in \row
          {
              \ifnum\cell=1
                  \draw[arc] (p\r) edge (p\c);
              \fi
              \ifnum\cell=6
                  \draw[arc,red] (p\r) edge (p\c);
              \fi
          }
      }
      \end{tikzpicture}
      }
      \resizebox{.23\textwidth}{!}{
      \begin{tikzpicture}[
      vertex/.style={draw,circle,minimum size=.3cm},
      arc/.style={draw,bend left=10,-{Latex[length=2mm, width=1mm]}}]
      \node[vertex] (p1) at (.62,.78) {};
      \node[vertex] (p2) at (-.22,.97) {};
      \node[vertex] (p3) at (-.9,.43) {};
      \node[vertex] (p4) at (-.9,-.43) {};
      \node[vertex] (p5) at (-.22,-.97) {}; 
      \node[vertex] (p6) at (.62,-.78) {};
      \node[vertex] (p7) at (1,0) {}; 
      \node[] (l1) at (1.5,0) {$+$}; 
      \foreach [count=\r] \row in 
      {{0,1,2,3,4,5,6},
       {6,0,1,2,3,4,5},
       {5,6,0,1,2,3,4},
       {4,5,6,0,1,2,3},
       {3,4,5,6,0,1,2},
       {2,3,4,5,6,0,1},
       {1,2,3,4,5,6,0}}
      {
          \foreach [count=\c] \cell in \row
          {
              \ifnum\cell=1
                  \draw[arc,Gray!40] (p\r) edge (p\c);
              \fi
              \ifnum\cell=3
                  \draw[arc,Gray!40] (p\r) edge (p\c);
              \fi
              \ifnum\cell=4
                  \draw[arc,Gray!40] (p\r) edge (p\c);
              \fi
              \ifnum\cell=6
                  \draw[arc,Gray!40] (p\r) edge (p\c);
              \fi
          }
      }
      \foreach [count=\r] \row in 
      {{0,1,2,3,4,5,6},
       {6,0,1,2,3,4,5},
       {5,6,0,1,2,3,4},
       {4,5,6,0,1,2,3},
       {3,4,5,6,0,1,2},
       {2,3,4,5,6,0,1},
       {1,2,3,4,5,6,0}}
      {
          \foreach [count=\c] \cell in \row
          {
              \ifnum\cell=2
                  \draw[arc,cyan] (p\r) edge (p\c);
              \fi
              \ifnum\cell=5
                  \draw[arc,violet] (p\r) edge (p\c);
              \fi
          }
      }
      \end{tikzpicture}
      }
      \resizebox{.188\textwidth}{!}{
      \begin{tikzpicture}[
      vertex/.style={draw,circle,minimum size=.3cm},
      arc/.style={draw,bend left=10,-{Latex[length=2mm, width=1mm]}}]
      \node[vertex] (p1) at (.62,.78) {};
      \node[vertex] (p2) at (-.22,.97) {};
      \node[vertex] (p3) at (-.9,.43) {};
      \node[vertex] (p4) at (-.9,-.43) {};
      \node[vertex] (p5) at (-.22,-.97) {}; 
      \node[vertex] (p6) at (.62,-.78) {};
      \node[vertex] (p7) at (1,0) {};  
      \foreach [count=\r] \row in 
      {{0,1,2,3,4,5,6},
       {6,0,1,2,3,4,5},
       {5,6,0,1,2,3,4},
       {4,5,6,0,1,2,3},
       {3,4,5,6,0,1,2},
       {2,3,4,5,6,0,1},
       {1,2,3,4,5,6,0}}
      {
          \foreach [count=\c] \cell in \row
          {
              \ifnum\cell=1
                  \draw[arc,Gray!40] (p\r) edge (p\c);
              \fi
              \ifnum\cell=2
                  \draw[arc,Gray!40] (p\r) edge (p\c);
              \fi
              \ifnum\cell=5
                  \draw[arc,Gray!40] (p\r) edge (p\c);
              \fi
              \ifnum\cell=6
                  \draw[arc,Gray!40] (p\r) edge (p\c);
              \fi
          }
      }
      \foreach [count=\r] \row in 
      {{0,1,2,3,4,5,6},
       {6,0,1,2,3,4,5},
       {5,6,0,1,2,3,4},
       {4,5,6,0,1,2,3},
       {3,4,5,6,0,1,2},
       {2,3,4,5,6,0,1},
       {1,2,3,4,5,6,0}}
      {
          \foreach [count=\c] \cell in \row
          {
              \ifnum\cell=3
                  \draw[arc,Orange] (p\r) edge (p\c);
              \fi
              \ifnum\cell=4
                  \draw[arc,ForestGreen] (p\r) edge (p\c);
              \fi
          }
      }
      \end{tikzpicture}
      } \\ \hline
      $N=8$ &
      \vspace{1mm}
      \resizebox{.23\textwidth}{!}{
      \begin{tikzpicture}[
      vertex/.style={draw,circle,minimum size=.3cm},
      arc/.style={draw,-{Latex[length=2mm, width=1mm]},bend left=10}]
      \node[vertex] (p1) at (.7,.7) {};
      \node[vertex] (p2) at (0,1) {};
      \node[vertex] (p3) at (-.7,.7) {};
      \node[vertex] (p4) at (-1,0) {};
      \node[vertex] (p5) at (-.7,-.7) {}; 
      \node[vertex] (p6) at (0,-1) {};
      \node[vertex] (p7) at (.7,-.7) {};
      \node[vertex] (p8) at (1,0) {};
      \node[] (l1) at (1.5,0) {$=$};
      \foreach [count=\r] \row in 
      {{0,1,3,8,5,6,7,2},
       {2,0,1,3,4,5,9,7},
       {7,2,0,1,3,8,5,6},
       {9,7,2,0,1,3,4,5},
       {5,6,7,2,0,1,3,8},
       {4,5,9,7,2,0,1,3},
       {3,8,5,6,7,2,0,1},
       {1,3,4,5,9,7,2,0}}
      {
          \foreach [count=\c] \cell in \row
          {
              \ifnum\cell=1
                  \draw[arc] (p\r) edge (p\c);
              \fi
              \ifnum\cell=2
                  \draw[arc,red] (p\r) edge (p\c);
              \fi
              \ifnum\cell=3
                  \draw[arc,cyan] (p\r) edge (p\c);
              \fi
              \ifnum\cell=4
                  \draw[arc,ForestGreen,bend left=0] (p\r) edge (p\c);
              \fi
              \ifnum\cell=5
                  \draw[arc,SpringGreen,bend left=0] (p\r) edge (p\c);
              \fi
              \ifnum\cell=6
                  \draw[arc,ForestGreen,bend left=0] (p\r) edge (p\c);
              \fi
              \ifnum\cell=7
                  \draw[arc,violet] (p\r) edge (p\c);
              \fi
              \ifnum\cell=8
                  \draw[arc,orange,bend left=0] (p\r) edge (p\c);
              \fi
              \ifnum\cell=9
                  \draw[arc,orange,bend left=0] (p\r) edge (p\c);
              \fi
          }
      }
      \end{tikzpicture}
      } \newline and 11 others~~~~~~~
      &
      \resizebox{.17\textwidth}{!}{
      \begin{tikzpicture}[
      vertex/.style={draw,circle,minimum size=.3cm},
      arc/.style={draw,-{Latex[length=2mm, width=1mm]},bend left=10}]
      \node[vertex] (p1) at (.7,.7) {};
      \node[vertex] (p2) at (0,1) {};
      \node[vertex] (p3) at (-.7,.7) {};
      \node[vertex] (p4) at (-1,0) {};
      \node[vertex] (p5) at (-.7,-.7) {}; 
      \node[vertex] (p6) at (0,-1) {};
      \node[vertex] (p7) at (.7,-.7) {};
      \node[vertex] (p8) at (1,0) {};
      \node[] (l1) at (1.5,0) {$+$};
      \foreach [count=\r] \row in 
      {{0,1,3,4,5,6,7,2},
       {2,0,1,3,4,5,6,7},
       {7,2,0,1,3,4,5,6},
       {6,7,2,0,1,3,4,5},
       {5,6,7,2,0,1,3,4},
       {4,5,6,7,2,0,1,3},
       {3,4,5,6,7,2,0,1},
       {1,3,4,5,6,7,2,0}}
      {
          \foreach [count=\c] \cell in \row
          {
              \ifnum\cell=3
                  \draw[arc,Gray!40] (p\r) edge (p\c);
              \fi
              \ifnum\cell=4
                  \draw[arc,Gray!40] (p\r) edge (p\c);
              \fi
              \ifnum\cell=6
                  \draw[arc,Gray!40] (p\r) edge (p\c);
              \fi
              \ifnum\cell=7
                  \draw[arc,Gray!40] (p\r) edge (p\c);
              \fi
          }
      }
      \foreach [count=\r] \row in 
      {{0,1,3,4,5,6,7,2},
       {2,0,1,3,4,5,6,7},
       {7,2,0,1,3,4,5,6},
       {6,7,2,0,1,3,4,5},
       {5,6,7,2,0,1,3,4},
       {4,5,6,7,2,0,1,3},
       {3,4,5,6,7,2,0,1},
       {1,3,4,5,6,7,2,0}}
      {
          \foreach [count=\c] \cell in \row
          {
              \ifnum\cell=1
                  \draw[arc] (p\r) edge (p\c);
              \fi
              \ifnum\cell=2
                  \draw[arc,red] (p\r) edge (p\c);
              \fi
              \ifnum\cell=5
                  \draw[arc,SpringGreen,bend left=0] (p\r) edge (p\c);
              \fi
          }
      }
      \end{tikzpicture}
      }
      \resizebox{.17\textwidth}{!}{
      \begin{tikzpicture}[
      vertex/.style={draw,circle,minimum size=.3cm},
      arc/.style={draw,-{Latex[length=2mm, width=1mm]},bend left=10}]
      \node[vertex] (p1) at (.7,.7) {};
      \node[vertex] (p2) at (0,1) {};
      \node[vertex] (p3) at (-.7,.7) {};
      \node[vertex] (p4) at (-1,0) {};
      \node[vertex] (p5) at (-.7,-.7) {}; 
      \node[vertex] (p6) at (0,-1) {};
      \node[vertex] (p7) at (.7,-.7) {};
      \node[vertex] (p8) at (1,0) {};
      \node[] (l1) at (1.5,0) {$+$};
      \foreach [count=\r] \row in 
      {{0,1,3,4,5,6,7,2},
       {2,0,1,3,4,5,6,7},
       {7,2,0,1,3,4,5,6},
       {6,7,2,0,1,3,4,5},
       {5,6,7,2,0,1,3,4},
       {4,5,6,7,2,0,1,3},
       {3,4,5,6,7,2,0,1},
       {1,3,4,5,6,7,2,0}}
      {
          \foreach [count=\c] \cell in \row
          {
              \ifnum\cell=1
                  \draw[arc,Gray!40] (p\r) edge (p\c);
              \fi
              \ifnum\cell=2
                  \draw[arc,Gray!40] (p\r) edge (p\c);
              \fi
              \ifnum\cell=4
                  \draw[arc,Gray!40] (p\r) edge (p\c);
              \fi
              \ifnum\cell=5
                  \draw[arc,Gray!40,bend left=0] (p\r) edge (p\c);
              \fi
              \ifnum\cell=6
                  \draw[arc,Gray!40] (p\r) edge (p\c);
              \fi
          }
      }
      \foreach [count=\r] \row in 
      {{0,1,3,4,5,6,7,2},
       {2,0,1,3,4,5,6,7},
       {7,2,0,1,3,4,5,6},
       {6,7,2,0,1,3,4,5},
       {5,6,7,2,0,1,3,4},
       {4,5,6,7,2,0,1,3},
       {3,4,5,6,7,2,0,1},
       {1,3,4,5,6,7,2,0}}
      {
          \foreach [count=\c] \cell in \row
          {
              \ifnum\cell=3
                  \draw[arc,cyan] (p\r) edge (p\c);
              \fi
              \ifnum\cell=7
                  \draw[arc,violet] (p\r) edge (p\c);
              \fi
          }
      }
      \end{tikzpicture}
      }
      \resizebox{.17\textwidth}{!}{
      \begin{tikzpicture}[
      vertex/.style={draw,circle,minimum size=.3cm},
      arc/.style={draw,-{Latex[length=2mm, width=1mm]},bend left=10}]
      \node[vertex] (p1) at (.7,.7) {};
      \node[vertex] (p2) at (0,1) {};
      \node[vertex] (p3) at (-.7,.7) {};
      \node[vertex] (p4) at (-1,0) {};
      \node[vertex] (p5) at (-.7,-.7) {}; 
      \node[vertex] (p6) at (0,-1) {};
      \node[vertex] (p7) at (.7,-.7) {};
      \node[vertex] (p8) at (1,0) {};
      \node[] (l1) at (1.5,0) {$+$};
      \foreach [count=\r] \row in 
      {{0,1,3,8,5,6,7,2},
       {2,0,1,3,4,5,9,7},
       {7,2,0,1,3,8,5,6},
       {9,7,2,0,1,3,4,5},
       {5,6,7,2,0,1,3,8},
       {4,5,9,7,2,0,1,3},
       {3,8,5,6,7,2,0,1},
       {1,3,4,5,9,7,2,0}}
      {
          \foreach [count=\c] \cell in \row
          {
              \ifnum\cell=1
                  \draw[arc,Gray!40] (p\r) edge (p\c);
              \fi
              \ifnum\cell=2
                  \draw[arc,Gray!40] (p\r) edge (p\c);
              \fi
              \ifnum\cell=3
                  \draw[arc,Gray!40] (p\r) edge (p\c);
              \fi
              \ifnum\cell=5
                  \draw[arc,Gray!40,bend left=0] (p\r) edge (p\c);
              \fi
              \ifnum\cell=7
                  \draw[arc,Gray!40] (p\r) edge (p\c);
              \fi
              \ifnum\cell=8
                  \draw[arc,Gray!40] (p\r) edge (p\c);
              \fi
              \ifnum\cell=9
                  \draw[arc,Gray!40] (p\r) edge (p\c);
              \fi
          }
      }
      \foreach [count=\r] \row in 
      {{0,1,3,8,5,6,7,2},
       {2,0,1,3,4,5,9,7},
       {7,2,0,1,3,8,5,6},
       {9,7,2,0,1,3,4,5},
       {5,6,7,2,0,1,3,8},
       {4,5,9,7,2,0,1,3},
       {3,8,5,6,7,2,0,1},
       {1,3,4,5,9,7,2,0}}
      {
          \foreach [count=\c] \cell in \row
          {
              \ifnum\cell=4
                  \draw[arc,ForestGreen,bend left=0] (p\r) edge (p\c);
              \fi
              \ifnum\cell=6
                  \draw[arc,ForestGreen,bend left=0] (p\r) edge (p\c);
              \fi
          }
      }
      \end{tikzpicture}
      }
    \resizebox{.14\textwidth}{!}{
      \begin{tikzpicture}[
      vertex/.style={draw,circle,minimum size=.3cm},
      arc/.style={draw,-{Latex[length=2mm, width=1mm]},bend left=10}]
      \node[vertex] (p1) at (.7,.7) {};
      \node[vertex] (p2) at (0,1) {};
      \node[vertex] (p3) at (-.7,.7) {};
      \node[vertex] (p4) at (-1,0) {};
      \node[vertex] (p5) at (-.7,-.7) {}; 
      \node[vertex] (p6) at (0,-1) {};
      \node[vertex] (p7) at (.7,-.7) {};
      \node[vertex] (p8) at (1,0) {};
      \foreach [count=\r] \row in 
      {{0,1,3,8,5,6,7,2},
       {2,0,1,3,4,5,9,7},
       {7,2,0,1,3,8,5,6},
       {9,7,2,0,1,3,4,5},
       {5,6,7,2,0,1,3,8},
       {4,5,9,7,2,0,1,3},
       {3,8,5,6,7,2,0,1},
       {1,3,4,5,9,7,2,0}}
      {
          \foreach [count=\c] \cell in \row
          {
              \ifnum\cell=1
                  \draw[arc,Gray!40] (p\r) edge (p\c);
              \fi
              \ifnum\cell=2
                  \draw[arc,Gray!40] (p\r) edge (p\c);
              \fi
              \ifnum\cell=3
                  \draw[arc,Gray!40] (p\r) edge (p\c);
              \fi
              \ifnum\cell=4
                  \draw[arc,Gray!40] (p\r) edge (p\c);
              \fi
              \ifnum\cell=5
                  \draw[arc,Gray!40,bend left=0] (p\r) edge (p\c);
              \fi
              \ifnum\cell=6
                  \draw[arc,Gray!40] (p\r) edge (p\c);
              \fi
              \ifnum\cell=7
                  \draw[arc,Gray!40] (p\r) edge (p\c);
              \fi
          }
      }
      \foreach [count=\r] \row in 
      {{0,1,3,8,5,6,7,2},
       {2,0,1,3,4,5,9,7},
       {7,2,0,1,3,8,5,6},
       {9,7,2,0,1,3,4,5},
       {5,6,7,2,0,1,3,8},
       {4,5,9,7,2,0,1,3},
       {3,8,5,6,7,2,0,1},
       {1,3,4,5,9,7,2,0}}
      {
          \foreach [count=\c] \cell in \row
          {
              \ifnum\cell=8
                  \draw[arc,orange,bend left=0] (p\r) edge (p\c);
              \fi
              \ifnum\cell=9
                  \draw[arc,orange,bend left=0] (p\r) edge (p\c);
              \fi
          }
      }
      \end{tikzpicture}
      }\\ \hline
    \end{tabular}}
\label{tbl:tbls1}
\end{center}
\end{table*}

Using this {\it AISync} strength $r$, we first enumerate all networks of a given size supporting {\it AISync} (Table~I).
For each $N$, we generate one or more diagrams representing all $N$-node symmetric networks, which are shown in the first row of Table~I for $N=3,4,5$. 
In these diagrams, each color indicates a set of links that, in any given symmetric network, must all exist together and be of the same type or not exist at all (noting that links from different sets can be of the same type).
For example, there are three distinct symmetric networks for $N=3$: a directed ring (cyan or black links),
an undirected ring (cyan and black links of the same type), 
and the superposition of two directed rings in opposite directions (cyan and black links of different types; as in Fig.~\ref{fig:2}). 
For a given symmetric network derived from these diagrams, we choose the external sublink pattern for each link type from all 
possible ways of connecting a subnode pair to another.
For the internal sublink patterns, we use either the binary or quaternary choices, where each node has one directed sublink (in either direction) in the binary case, while all four possibilities are allowed in the quaternary case.
The rest of Table~I lists the total numbers of isomorphically distinct external sublink structures with $r>0.05$, $r>0.2$, and (optimal) $r=1$; see Supplementary Tables~S1 and S2 for all optimal networks with $N=3$ and $4$, respectively.

{\YZ
Table~II extends the first row in Table~I, showing the symmetric network diagrams for $N=6,7$, and 8.
In each row, the leftmost diagram is the full representation as in Table~I, which is decomposed into multiple components (the partial diagrams in the same row) to make them more clearly visible.
The partial diagrams with the same background color indicate identical components appearing in multiple rows.
Thus, for $N=6$, we have four different diagrams (rows), each with a different combination of components.
There is only one diagram for $N=7$, while we show one representative diagram out of twelve in the case of $N=8$.
}

\begin{figure}
\centering
\includegraphics[width=.8\columnwidth]{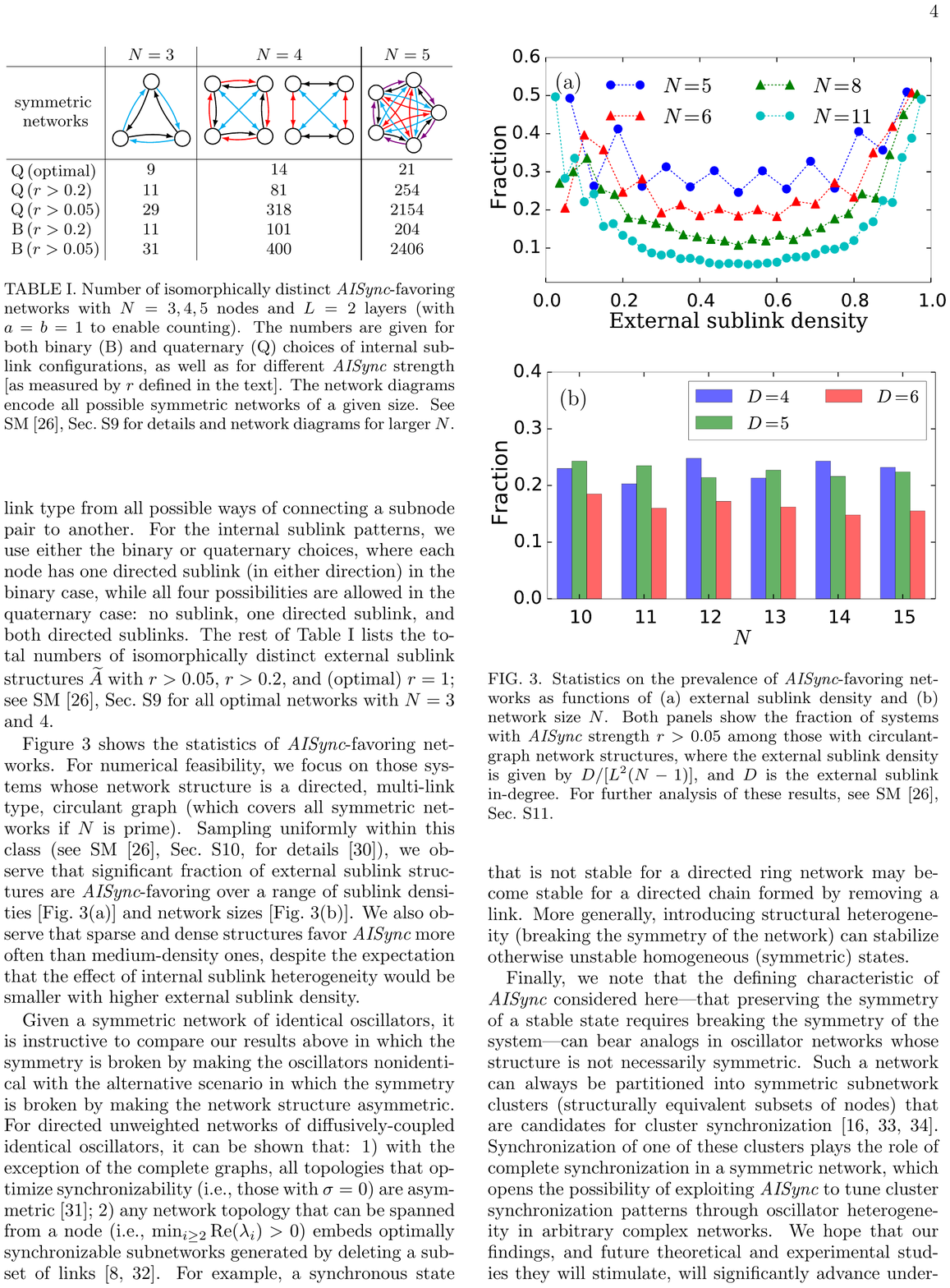}
\caption{
Statistics on the prevalence of \textit{AISync}-favoring networks.
Shown as functions of (a) external sublink density and (b) network size $N$.
Both panels show the fraction of systems with {\it AISync} strength $r>0.05$ among those with circulant network structures, where the external sublink density is given by $D/[L^2(N-1)]$, and $D$ is the number of external sublinks received by a node (which is the same for all nodes).
}
\label{fig:4}
\end{figure}

Figure~\ref{fig:4} shows the statistics of {\it AISync}-favoring networks.
For numerical feasibility, we focus on those systems whose network structure is a directed circulant graph with multiple link types (which covers all symmetric networks if $N$ is a prime number).
Sampling uniformly within this class (Appendix~\ref{sec:s4}), 
we observe that significant fraction of external sublink structures are {\it AISync}-favoring over a range of external sublink densities [Fig.~\ref{fig:4}(a)] and network sizes [Fig.~\ref{fig:4}(b)].
We also observe that sparse and dense structures favor {\it AISync} more often than medium-density ones, despite the expectation that the effect of internal sublink heterogeneity would be smaller with higher external sublink density. 
This phenomenon is further explored in Appendix~\ref{sec:symmetry-fig3} by establishing the approximate left-right symmetry in Fig.~\ref{fig:4}(a).

\section{Discussion}

Given a symmetric network of identical oscillators, it is instructive to compare our results above in which the symmetry is broken by making the oscillators nonidentical with the alternative scenario in which the symmetry is broken by making the network structure asymmetric. For directed unweighted  networks of diffusively-coupled identical oscillators, it can be shown that: 1) with the exception of the complete graphs, all topologies that optimize synchronizability (i.e., those with $\sigma=0$) are asymmetric; 2) any 
network topology that can be spanned from a node (i.e., $\min_{i\ge2}\text{Re}(\lambda_i) > 0$)
embeds optimally synchronizable subnetworks generated by deleting 
a subset of links~\cite{nishikawa2006synchronization,nishikawa2010network}. 
For example, a synchronous state that is not stable for a directed ring network may become stable for a directed chain formed by removing a link. 
More generally, introducing structural heterogeneity (breaking the symmetry of the network) can stabilize otherwise unstable homogeneous (symmetric) states.
 
Finally, we note that the defining characteristic of {\it AISync} considered here---that preserving the symmetry of a stable state requires breaking the symmetry of the system---can bear analogs in oscillator networks whose structure is not necessarily symmetric.
Such a network can always be partitioned into symmetric subnetwork clusters (structurally equivalent subsets of nodes) that are candidates for cluster synchronization~\cite{golubitsky2016rigid,pecora2014cluster,nicosia2013remote}.
Synchronization of one of these clusters plays the role of complete synchronization in a symmetric network, which opens the possibility of exploiting {\it AISync} to tune cluster synchronization patterns through oscillator heterogeneity in arbitrary complex networks.
We hope that our findings, and future theoretical and experimental studies they will stimulate, will significantly advance understanding of the interplay between symmetry and network dynamics.

\begin{acknowledgments}
The authors thank Thomas Wytock and Young Sul Cho for insightful discussions.
This work was supported by ARO Grant No. W911NF-15-1-0272.
\end{acknowledgments}

\appendix

\section{Definition of Cayley graphs}
\label{sec:cayley}

Given a generating set $S$ of a finite group $G$, the Cayley graph associated with $S$ and $G$ is defined as the network in which a node represents an element of $G$ and a directed link from one node $g\in G$ to another $g'\in G$ represents the composition of some element $s \in S$ with $g$ that gives $g'$ (i.e., $gs=g'$). While such a network is generally directed, it will be undirected if the inverse of every element of $S$ belongs to $S$. 
Choosing $S$ to be a generating set guarantees that the resulting network is (strongly) connected.
A generalization to multiple link types can be obtained if we assign different elements of $S$ to different link types.

\section{Details on multilayer models}
\label{sec:mapping}

Since Eq.~\eqref{eq:multilayer} defines a subclass of systems governed by Eq.~\eqref{eq:0}, it can always be written in the form of Eq.~\eqref{eq:0} for a given network structure specified by $A^{(\alpha)}$. This can be seen by stacking the $m$-dimensional vectors in Eq.~\eqref{eq:multilayer} and defining appropriate functions as follows:
\begin{gather}
\bm{X}_i := \begin{pmatrix}
\bm{x}^{(i)}_1\\
\vdots\\
\bm{x}^{(i)}_L
\end{pmatrix},\quad
\bm{F}_i := \begin{pmatrix}
\bm{F}^{(i)}_1\\
\vdots\\
\bm{F}^{(i)}_L
\end{pmatrix},\quad
\bm{H}^{(\alpha)} := \begin{pmatrix}
\bm{H}^{(\alpha)}_1\\
\vdots\\
\bm{H}^{(\alpha)}_L
\end{pmatrix},\\
\bm{F}^{(i)}_{\ell}(\bm{X}_i) := 
\bm{f}(\bm{x}^{(i)}_{\ell}) + \sum_{\ell'=1}^{L} \widetilde{A}^{(ii)}_{\ell\ell'} [\bm{h}(\bm{x}^{(i)}_{\ell'}) - \bm{h}(\bm{x}^{(i)}_{\ell})],\\
\bm{H}^{(\alpha)}_{\ell}(\bm{X}_i, \bm{X}_{i'})
:= \sum_{\ell'=1}^{L} B^{(\alpha)}_{\ell\ell'} [\bm{h}(\bm{x}^{(i')}_{\ell'}) - \bm{h}(\bm{x}^{(i)}_{\ell})],
\end{gather}
where $B^{(\alpha)}_{\ell\ell'}$ is defined to be the value of $\widetilde{A}^{(ii')}_{\ell\ell'}$ when node $i'$ is connected to node $i$ by a link of type $\alpha$.
Note that these node-to-node interactions are not necessarily diffusive, since we can have $\bm{H}^{(\alpha)}(\bm{X}_i, \bm{X}_{i'})\neq\bm{0}$ even for $\bm{X}_i=\bm{X}_{i'}$, if $\bm{x}^{(i)}_{\ell} \neq \bm{x}^{(i')}_{\ell'}$ for some $\ell\neq\ell'$
{\TN
[which in particular means that the coupling term cannot be written in the form ${\bm H}^{(\alpha)}({\bm X}_i, {\bm X}_{i'}) = \widetilde{\bm H}^{(\alpha)}({\bm X}_{i'}) - \widetilde{\bm H}^{(\alpha)}({\bm X}_i)$].  For example, even when nodes $1$ and $4$ are synchronized in the network of Fig. 1, i.e., ${\bm X}_1 = {\bm X}_4 = (\bm{s}_1(t), \bm{s}_2(t))^T$, the coupling term corresponding to the link of type $\alpha=3$ between them is in general not identically zero:
\begin{equation}
    \bm{H}^{(3)}(\bm{X}_1,\bm{X}_4) 
    = \begin{pmatrix}
    \bm{h}(\bm{s}_2) - \bm{h}(\bm{s}_1)\\
    0
    \end{pmatrix}
    \not\equiv \bm{0}.
\end{equation}
}
However, since we assume identical dynamics for subnodes and diffusive coupling between subnodes, a synchronous state of Eq.~\eqref{eq:multilayer} given by $\bm{x}^{(i)}_{\ell}(t)=\bm{s}(t)$, $\forall i,\ell$ with $\dot{\bm{s}} = \bm{f}\left( \bm{s} \right)$ is guaranteed to exist even if $F^{(i)}$'s are heterogeneous.
This corresponds to a global synchronous state of Eq.~\eqref{eq:0} defined by $\bm{X}_i = \bm{S} := (\bm{s},\ldots,\bm{s})^T$, $\forall i$, which can be verified by noting that $\bm{H}^{(\alpha)}(\bm{S}, \bm{S}) = \bm{0}$ and $\bm{F}^{(i)}_{\ell}(\bm{S}) := \bm{f}(\bm{s})$, $\forall i,\ell$.

\section{Details on MSF analysis}
\label{sec:msf}

Equation~\eqref{eq:multilayer} can be rewritten as a monolayer network by defining a single index for all the $n:=LN$ subnodes, in which node $i$ has subnodes $j=k_{i1},\ldots,k_{iL}$ with $k_{i\ell} :=(i-1)L+\ell$.
This leads to the standard form for a (monolayer) diffusively coupled network of oscillators:
\begin{equation}
\dot{\bm{x}}_j = \bm{f}(\bm{x}_j) + \sum_{j'=1}^n \widetilde{A}_{jj'} [\bm{h}(\bm{x}_{j'}) - \bm{h}(\bm{x}_{j})],
\label{eq:1}
\end{equation}
where $\bm{x}_j=\bm{x}^{(i)}_{\ell}$ and $\widetilde{A}_{jj'}:=\widetilde{A}^{(ii')}_{\ell\ell'}$ for $j=k_{i\ell}$ and $j'=k_{i'\ell'}$.
In the monolayer adjacency matrix $\widetilde{A} = (\widetilde{A}_{jj'})$, the matrix $B^{(\alpha)} = (B^{(\alpha)}_{\ell\ell'})$ appears as multiple off-diagonal blocks of size $L$, and the arrangement of those blocks within $\widetilde{A}$ matches with the structure of the corresponding adjacency matrix $A^{(\alpha)}$, reflecting the topology of node-to-node interactions through links of type $\alpha$ [see Fig.~\ref{fig:1}(d) for an example].
This equation allows application of the MSF analysis~\cite{pecora1998master} because subnodes and sublinks (and the associated coupling functions) are identical.
The stability function $\psi(\lambda)$ is defined as the maximum Lyapunov exponent of the reduced variational equation,
\begin{equation}
\dot{\bm{\xi}} = [D\bm{f}(\bm{s}) - \lambda D\bm{h}(\bm{s})] \bm{\xi},
\label{eqn:mse}
\end{equation}
where $\bm{\xi}$ is an $m$-dimensional perturbation vector, $D\bm{f}(\bm{s})$ and $D\bm{h}(\bm{s})$ are the Jacobian of $\bm{f}$ and $\bm{h}$, respectively, at the synchronous state, $\bm{x}_j=\bm{s}(t)$, $\forall j$, and $\lambda$ is a complex-valued parameter.

\section{Verifying the \textit{AIS\MakeLowercase{ync}} conditions}
\label{sec:s2}

Here we describe our scheme for verifying {\it AISync} conditions (C1) and (C2) given a symmetric network structure (adjacency matrices $A^{(\alpha)}$), external sublink configurations (matrices $B^{(\alpha)}$), a set $\mathcal{F}$ of possible internal sublink configurations (from which matrices $F^{(i)}$ are chosen), isolated subnode dynamics $\bm{f}$, and sublink coupling function $\bm{h}$.
We first obtain the stability function $\psi(\lambda)$: 
\begin{enumerate}
\item Compute a trajectory $\bm{s}(t)$ of an isolated subnode by integrating $\dot{\bm{s}} = \bm{f}\left( \bm{s} \right)$, which determines the synchronous state, $\bm{x}^{(i)}_{\ell}(t)=\bm{s}(t)$, $\forall i,\ell$.
\item Integrate Eq.~\eqref{eqn:mse} and calculate its maximum Lyapunov exponent (MLE), which defines $\psi(\lambda)$ for a range of $\lambda$ in the complex plane.
\end{enumerate}
Note that $\psi(\lambda)$ depends only on $\bm{f}$, $\bm{h}$, and $\bm{s}(t)$.
For a given symmetric network structure and external sublink configurations, we can compute the stability $\Psi$ of the synchronous state for any combination of $F^{(i)}\in\mathcal{F}$ by calculating and substituting the Laplacian eigenvalues $\lambda_j$ into the formula $\Psi = \max_{2\leq j \leq n}\psi(\lambda_j)$.
To establish the {\it AISync} property, we verify the following conditions:
\begin{itemize}
\item (C1)$'$: For each matrix $F \in \mathcal{F}$, set $F^{(1)} = \cdots = F^{(N)} = F$ (leading to a homogeneous system) and verify $\Psi>0$.
\item (C2)$'$: Identify a combination of (heterogeneous) $F^{(i)}\in\mathcal{F}$ for which $\Psi<0$ (e.g., by checking exhaustively or by using a numerical optimization algorithm to minimize $\Psi$ over $F^{(i)}$).
\end{itemize}
The verification of condition (C1)$'$ provides strong support for (C1), since the only other possibility for a stable synchronization of all nodes is a state of the form $\bm{x}^{(i)}_{\ell}=\bm{s}_\ell(t)$, $\forall i, \ell$, with at least one $\bm{s}_\ell(t)$ different from the others (which we find does not exist in many cases, such as the examples in Fig.~\ref{fig:2} and in Supplemental Material Sec.~S1).
To provide additional support for (C1), we directly simulate Eq.~\eqref{eq:multilayer} from a set of initial conditions and verify that the synchronization error $e$ does not approach zero whenever $F^{(1)} = \cdots = F^{(N)}$, where $e$ is defined as the standard deviation of the node state vectors, or equivalently,
\begin{equation}
e^2 := \frac{1}{N}\sum_{i=1}^N \sum_{\ell=1}^L || \bm{x}_\ell^{(i)} - \overline{\bm{x}}_\ell ||^2, \quad
\overline{\bm{x}}_\ell := \frac{1}{N}\sum_{i=1}^N \bm{x}_\ell^{(i)}.
\label{eqn:error}
\end{equation} 
Here $||\cdot||$ denotes the $2$-norm in the state space of the subnode dynamics, and $e=0$ is achieved if and only if the system is in a synchronous state of the form $\bm{x}^{(i)}_{\ell}=\bm{s}_\ell(t)$.
To complete our procedure, we verify condition (C2)$'$, which rigorously establishes (C2).

\section{Details on example in Fig.~2}
\label{sec:msf-y-x-coupling}

In the example system from Fig.~\ref{fig:2}, the coupling matrices for the two link types are $B^{(1)} = (\begin{smallmatrix} b&b\\ 0&0 \end{smallmatrix})$ and $B^{(2)} = (\begin{smallmatrix} 0&0\\ 0&b \end{smallmatrix})$, where the constant $b$ represents the coupling strength common to all external sublinks.
The coupling matrix $\Fi$ for internal sublinks is chosen from the binary set $\mathcal{F}=\{ (\begin{smallmatrix} 0&a\\ 0&0 \end{smallmatrix}), (\begin{smallmatrix} 0&0\\ a&0 \end{smallmatrix}) \}$, corresponding to the two possible sublink directions [and thus to two types of nodes indicated by green and cyan color, respectively, in Fig.~\ref{fig:2}(a)], where the constant $a$ represents the coupling strength common to all internal sublinks.
The Lorenz dynamics of the subnodes and the coupling represented by sublinks are given by 
\begin{equation}
\begin{split}
\bm{f}(\bm{x}) = \begin{pmatrix}
\gamma(x_2 - x_1)\\
x_1(\rho - x_3) - x_2\\
x_1 x_2 - \beta x_3
\end{pmatrix}, \\
\bm{h}(\bm{x}) = \begin{pmatrix}x_2\\ 0\\ 0 \end{pmatrix}, \quad
\bm{x} = \begin{pmatrix}x_1\\ x_2\\ x_3 \end{pmatrix}
\label{eq:f-h-lorenz}
\end{split}
\end{equation}
with the standard parameters, $\gamma = 10$, $\rho = 28$, and $\beta = 8/3$.

\begin{figure}[t]
\centering
\includegraphics[width=1\columnwidth]{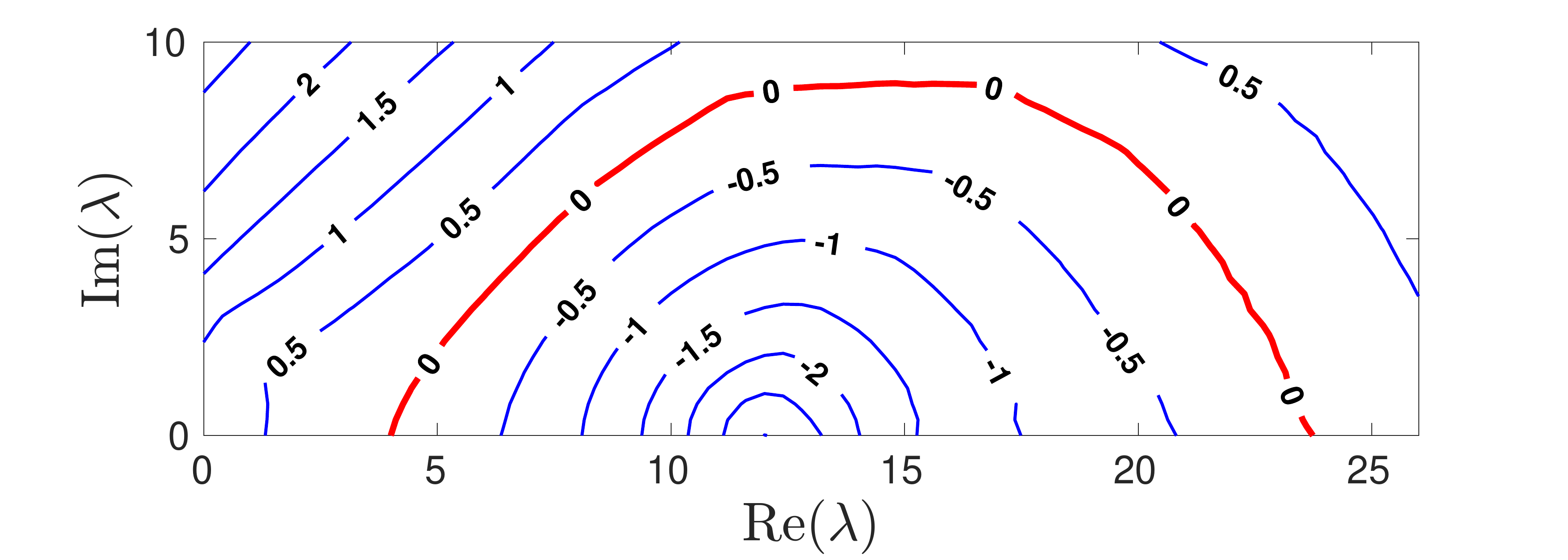}
\caption{Stability function $\psi(\lambda)$ for the {\it AISync} system in Fig.~\ref{fig:2}.}
\label{fig:s1}
\end{figure}

The stability function $\psi(\lambda)$ is determined by Eq.~\eqref{eqn:mse}, which for this system reads 
\begin{equation}
\begin{pmatrix} \dot{\xi}_1 \\ \dot{\xi}_2 \\ \dot{\xi}_3 \end{pmatrix} =
\begin{pmatrix}
-\gamma & \gamma - \lambda & 0 \\
\rho - s_3 & -1 & -s_1 \\
s_2 & s_1 & -\beta
\end{pmatrix}
\begin{pmatrix} \xi_1 \\ \xi_2 \\ \xi_3 \end{pmatrix},
\label{eq:mse-lorenz}
\end{equation}
where $\bm{\xi} := (\xi_1, \xi_2, \xi_3)^T$ is the variation of the state vector $\bm{x}$ and the synchronous state $\bm{s} := (s_1, s_2, s_3)^T$ satisfies the equation for a single isolated Lorenz oscillator:
\begin{equation}
\begin{split}
\dot{s}_1 &= \gamma(s_2 - s_1),\\
\dot{s}_2 &= s_1(\rho - s_3) - s_2,\\
\dot{s}_3 &= s_1 s_2 - \beta s_3.
\end{split}
\label{eqn:lorenz-s}
\end{equation}
For a given $\lambda$ in the complex plane, we compute $\psi(\lambda)$ by numerically integrating Eqs.~\eqref{eq:mse-lorenz} and \eqref{eqn:lorenz-s} for $2 \times 10^4$ time units and estimating the MLE~\cite{wolf1985determining} associated with the variable $\bm{\xi}$.
Figure~\ref{fig:s1} 
shows the resulting estimate of $\psi(\lambda)$, which has a bounded stability region $\{ \lambda\in\mathbb{C} \,\vert\, \psi(\lambda) < 0 \}$.

For a given combination of $a$ and $b$, we obtain $\Psi_{=}$ and $\Psi_{\neq}$, which are shown in Fig.~\ref{fig:2}(b). 
Note that for this example there are only two distinct homogeneous systems and two distinct heterogeneous systems. One of these heterogeneous systems is shown in Fig.~\ref{fig:2}(a).
We also note that $\Psi_{=} > 0$ and $\Psi_{\neq} < 0$ are equivalent to the conditions (C1)$'$ and (C2)$'$ in Appendix~\ref{sec:s2}, respectively.
For each combination of $a$ and $b$ satisfying both conditions [on a grid covering Fig.~\ref{fig:2}(b) with a resolution of $0.2$], we additionally run $24$ direct simulations of Eq.~\eqref{eq:multilayer} for $200$ time units.
The initial condition $\bm{x}^{(i)}_{\ell}(0)$ for each subnode is chosen randomly and independently from the uniform distribution in the region $[0,10]\times[0,10]\times[0,10]$ of its state space.
The results confirm that the synchronization error $e$ defined in Eq.~\eqref{eqn:error} and
averaged over the last $100$ time units does not fall below $10^{-3}$ in all $24$ runs for both homogeneous systems, providing solid evidence that the system satisfies the {\it AISync} condition (C1) for these combinations of $a$ and $b$.
Since $\Psi_{\neq} < 0$ implies (C2)$'$ and thus (C2), this confirms {\it AISync} in the region shaded purple in Fig.~\ref{fig:2}(b).

The initial condition for the sample trajectory in Fig.~\ref{fig:2}(c) is chosen randomly within a distance of $10^{-3}$ from the synchronous state.
The trajectory is then computed by integrating the system with all nodes green for $t\le25$, instantaneously switching the direction of the sublink between subnodes $2'$ and $2''$, and then continuing to integrate for $25\le t\le50$.




\section{Sampling protocol used in Fig.~\ref{fig:3}}
\label{sec:s4}

We randomly sample systems whose network structure $A^{(\alpha)}$ is 
a circulant graph (with directed links of possibly multiple types)
of given size $N$ and external sublink in-degree $D$ (i.e., the total number of sublinks received by the subnodes of a given node).
Each of the $D$ sublinks coming into node $1$ is chosen randomly; it connects a random subnode chosen uniformly from the other $N-1$ nodes to a random subnode chosen uniformly from node $1$.
The incoming sublinks into nodes $2$ to $N$ are then chosen to precisely match those coming into node $1$, which ensures that the network structure is 
a circulant graph.
This simultaneously specifies $A^{(\alpha)}$ and $B^{(\alpha)}$ defining the system.
To determine $\sigma_{\neq}$, $\sigma_{=}$, and $r$ for this system, we calculate the eigenspread $\sigma$ of the monolayer network representation for all the possible internal sublink configurations $\Fi$, chosen here from the binary set $\{ (\begin{smallmatrix} 0&1\\ 0&0 \end{smallmatrix}), (\begin{smallmatrix} 0&0\\ 1&0 \end{smallmatrix}) \}$.
For each combination of $N$ and $D$, we generate a sample of $4{,}000$ such systems to compute the fraction of {\it AISync}-favoring networks.

\section{Approximate symmetry in Fig.~\ref{fig:3}(\MakeLowercase{a})}
\label{sec:symmetry-fig3}

The approximate symmetry with respect to the vertical line at density $0.5$ observed in Fig.~\ref{fig:3}(\MakeLowercase{a}) can be explained using the notion of network complement.
The complement of a given (unweighted) network with adjacency matrix $\widetilde{A}=(\widetilde{A}_{jj'})$ is defined as the network having the adjacency matrix $\widetilde{A}^c=(\widetilde{A}_{jj'}^c)$ given by
\begin{equation}
\widetilde{A}_{jj'}^c := (1 - \widetilde{A}_{jj'})(1-\delta_{jj'}).
\end{equation}
The external sublink density of a network and its complement add up to one, placing them symmetrically about the vertical line at density $0.5$ in Fig.~\ref{fig:3}(\MakeLowercase{a}).
When the nontrivial Laplacian eigenvalues of the network and its complement, which we denote $\lambda_2,\ldots,\lambda_n$ and $\lambda_2^c,\ldots,\lambda_n^c$, respectively, are both indexed in the order of increasing real part, they are related by $\lambda_j + \lambda_{n+2-j}^c = n$~\cite{nishikawa2010network}.
This implies that, if $\sigma$ is the eigenvalue spread for a monolayer network with given internal sublink configurations $\Fi$, then the spread for its complement is given by 
\begin{equation}
\sigma^c = \frac{\widetilde{m} \sigma}{n(n-1) - \widetilde{m}},
\label{eqn:sigma-c}
\end{equation}
where $\widetilde{m} := \sum_j\sum_{j' \neq j} \widetilde{A}_{jj'}$ is the number of directed links in the network $\widetilde{A}$.
Now consider two systems with $n$ subnodes and adjacency matrices $\widetilde{A}_1$ and $\widetilde{A}_2$, whose $\sigma$ values are $\sigma_1$ and $\sigma_2$, respectively.
If we denote the $\sigma$ values of the complement of these systems by $\sigma_1^c$ and $\sigma_2^c$, respectively, we have
\begin{equation}
\frac{\sigma_1}{\sigma_2} = \frac{\sigma_1^c}{\sigma_2^c}
\end{equation}
when $\widetilde{A}_1$ and $\widetilde{A}_2$ have the same number of directed links, i.e., $\widetilde{m}_1=\widetilde{m}_2$.
It follows from Eq.~\eqref{eqn:sigma-c} that if $\widetilde{A}_1$ is the best homogeneous system and $\widetilde{A}_2$ the best heterogeneous system for a given external connection pattern with density $x$, then their complements are the best homogeneous and heterogeneous system for an external connection pattern with density $1-x$.
Thus, the value of {\it AISync} strength $r$ is the same at density $x$ and $1-x$. 
The symmetry, however, is not perfect between sparse and dense parts of the plot, since we exclude the cases in which the network is not synchronizable (i.e., we require $\min_{j\ge2}\text{Re}(\lambda_j) > 0$), the effect of which is not symmetric between sparse and dense cases.

\bibliography{net_dyn,net_str}

\clearpage
\onecolumngrid
\setcounter{page}{1}

\setcounter{equation}{0}
\renewcommand{\theequation}{S\arabic{equation}}
\setcounter{figure}{0}
\renewcommand{\thefigure}{S\arabic{figure}}
\setcounter{section}{0}
\renewcommand{\thesection}{S\arabic{section}}
\setcounter{table}{0}
\renewcommand{\thetable}{S\arabic{table}}

\begin{center}
{\Large\bf Supplemental Material}\\[3mm]
\textit{\mytitle}\\[1pt]
Yuanzhao Zhang, Takashi Nishikawa, and Adilson E. Motter
\end{center}

\section{Example of \textit{AIS\MakeLowercase{ync}} with unbounded stability region}
\label{sec:unbound-example}

\begin{figure*}[ht]
\centering
\includegraphics[width=.85\textwidth]{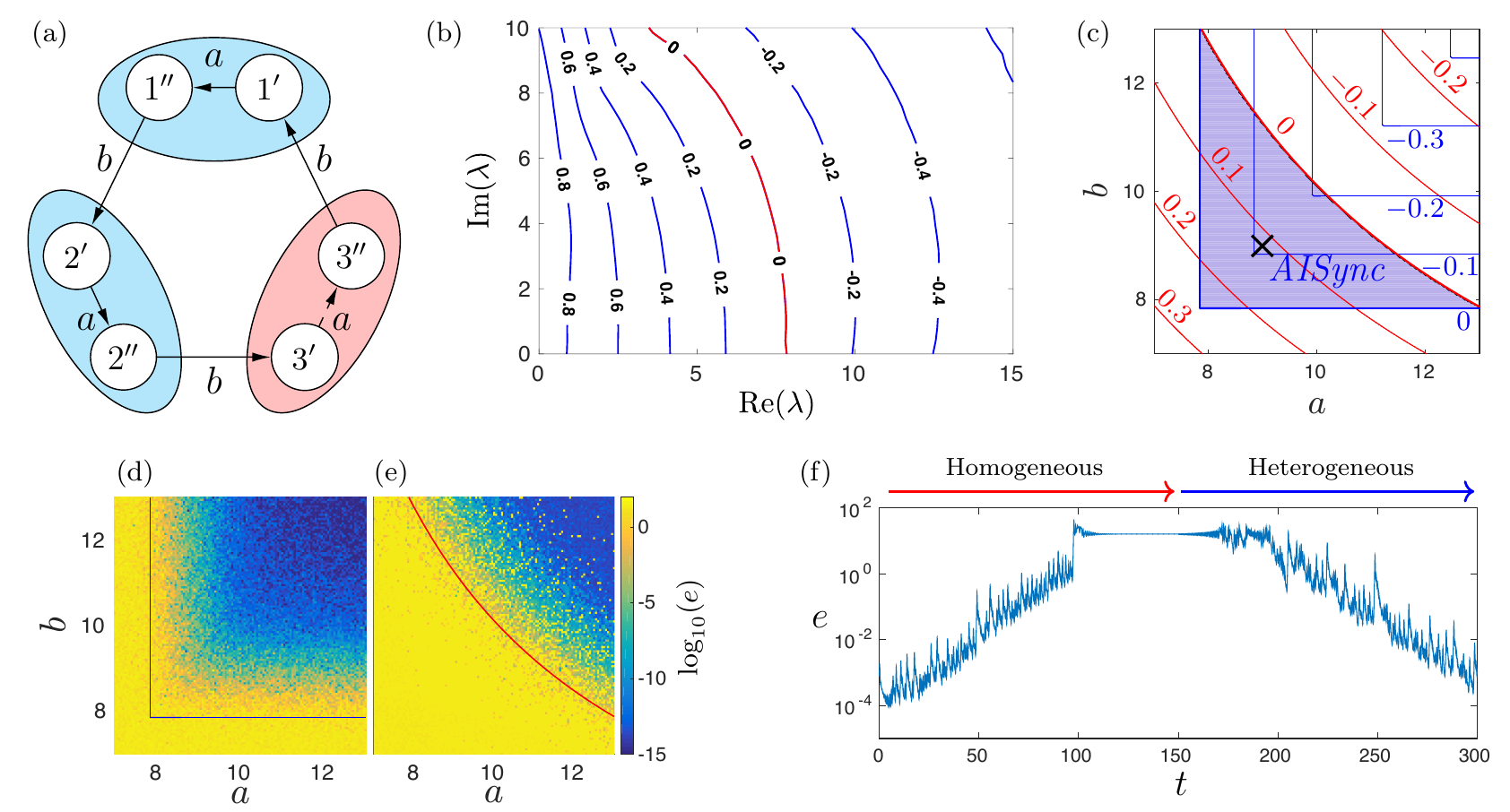}
\caption{Example of {\it AISync} systems that has unbounded stability region.
The subnodes are identical Lorenz oscillators, coupled to each other through the first component of their state vectors.
(a) Symmetric network of $N=3$ nodes, with $L=2$ subnodes per node, shown in the monolayer network representation for a specific heterogeneous system (when the dashed sublink does not exist).
If we add the dashed sublink, 
the system becomes homogeneous.
(b) Contour curves for the stability function $\psi(\lambda)$ in the complex plane.
The stability region consists of all points to the right of the red curve (corresponding to $\psi = 0$) and thus is unbounded.
(c) Contour curves for the MTLE for the homogeneous (red) and heterogeneous (blue) systems as a function of parameters $a$ and $b$.
The region of {\it AISync} is shaded light purple.
(d, e) Synchronization error $e$ after $200$ time units for the heterogeneous (d) and the homogeneous (e) system.
(f) Time evolution of the synchronization error $e$ for the system at $a=b=9$ [the cross symbol in (c)] undergoing a switch from being homogeneous to being heterogeneous at $t=150$ 
[through the removal of the internal sublink in node $3$].
}
\label{fig:s3}
\end{figure*}

Figure~\ref{fig:s3} shows another example of {\it AISync} using systems in which subnode dynamics is the same Lorenz oscillator as in Fig.~2, but with a different coupling function [leading to a different stability function $\psi(\lambda)$] and a different symmetric network.
We use a three-node symmetric network with a single link type [with the corresponding coupling matrix $B^{(1)} = (\begin{smallmatrix} 0&b\\ 0&0 \end{smallmatrix})$], as can be seen in Fig.~\ref{fig:s3}(a), where the monolayer representation of an example heterogenous system is shown.
The coupling matrix $\Fi$ for internal sublinks are chosen from the quaternary set $\{ (\begin{smallmatrix} 0&0\\ 0&0 \end{smallmatrix}), (\begin{smallmatrix} 0&a\\ 0&0 \end{smallmatrix}), (\begin{smallmatrix} 0&0\\ a&0 \end{smallmatrix}), (\begin{smallmatrix} 0&a\\ a&0 \end{smallmatrix}) \}$, so there are four distinct homogeneous systems. 
Two of them have $\lambda_2=0$ (and thus are not synchronizable), and one of the remaining ones 
is always more stable (i.e., smaller $\Psi$) than the other in the range of $a$ and $b$ considered in Fig.~\ref{fig:s3}.

The more stable
homogeneous system is the one with the dashed sublink in Fig.~\ref{fig:s3}(a).
Among the heterogeneous systems, the one with the smallest value of $\sigma$ is the one without the dashed sublink in Fig.~\ref{fig:s3}(a), which is optimal (i.e., $\sigma=0$).
We thus have just one homogeneous system and one heterogeneous system to compare. 
For the coupling function, we use $\bm{h}(\bm{x}) = (x_1, 0, 0)^T$, which leads to the stability function $\psi(\lambda)$ shown in Fig.~\ref{fig:s3}(b).
Unlike the example in Fig.~2, the stable region for this example (the region to the right of the red curve) is unbounded.
Using this $\psi(\lambda)$, we calculate the MTLEs for the homogeneous and heterogeneous systems, which are shown in Fig.~\ref{fig:s3}(c) as functions of parameters $a$ and $b$.
We confirm that the global stability also follows the same trend by integrating Eq.~(2) directly for $200$ time units from a random initial condition and then computing the synchronization error $e$ 
defined in Eq.~(D1).
The result is shown in Fig.~\ref{fig:s3}(d), where we see that $e$ behaves similarly as the MTLE. 
Figure~\ref{fig:s3}(e) shows the synchronization error $e$ for a typical trajectory of the system with $a=b=9$ [marked by black cross in Fig.~\ref{fig:s3}(c)], which switches from homogeneous to heterogeneous at $t=150$ with the removal the sublink inside the pink node in Fig.~\ref{fig:s3}(a).

\section{Example of \textit{AIS\MakeLowercase{ync}} in undirected networks}
\label{sec:undirected-networks}

The simplest example of undirected network exhibiting {\it AISync} is the network of two nodes, coupled bidirectionally.
Consider a system with this network structure in which node $1$ has subnodes $1'$ and $1''$, and node $2$ has subnodes $2'$ and $2''$, and the network of sublinks (all with weight equal to one) forms a directed chain connecting the four subnodes as $2'' \to 1' \to 1'' \to 2'$.
This network, when described at the node level, is indeed undirected because the pattern of sublink connections from node $1$ to $2$ is identical to the pattern of connections from node $2$ to $1$.
The nodes are heterogeneous because subnode $1'$ is connected to $1''$, while subnode $2'$ is not connected to $2''$.
Since the directed chain is an optimal network with $\lambda_2=\lambda_3=\lambda_4=1$, this heterogeneous system is more synchronizable than any combination of internal sublink configurations that leads to a homogeneous system.
Thus, this undirected network exhibits {\it AISync}.

\section{Supplementary Tables}

For each of the symmetric networks that can be derived from the network diagrams in Table~I for $N=3$ and $4$, we identify all possible two-layer optimal heterogeneous systems with that symmetric network structure.
Tables~S1 and S2 show all these systems in the monolayer representation for $N=3$ and $N=4$, respectively

\begin{table*}[ht]
\begin{center}
\caption{The $9$ optimal heterogeneous systems with $N=3$ and $L=2$.}
{\renewcommand{\arraystretch}{1.2}
\begin{tabular}{ | M{.16\textwidth} | m{.8\textwidth} | }

        \hline
        symmetric network & \hspace{2in}optimal heterogeneous systems\\
        \hline
        \resizebox{.15\textwidth}{!}{
        \begin{tikzpicture}[
        vertex/.style={draw,circle,very thick,minimum size=.6cm},
        arc/.style={draw,very thick,-{Latex[length=3mm, width=2mm]},bend left=10}]
        \node[vertex] (p1) at (0,2.18) {};
        \node[vertex] (p2) at (-1.18,0) {};
        \node[vertex] (p3) at (1.18,0) {};   
        \foreach [count=\r] \row in 
        {{0,1,2},
         {2,0,1},
         {1,2,0}}
        {
            \foreach [count=\c] \cell in \row
            {
                \ifnum\cell=2
                    \draw[arc,cyan] (p\r) edge (p\c);
                \fi
            }
        }
        \end{tikzpicture}
        } &
        \vspace{2mm}

        \includegraphics[width=.18\textwidth]{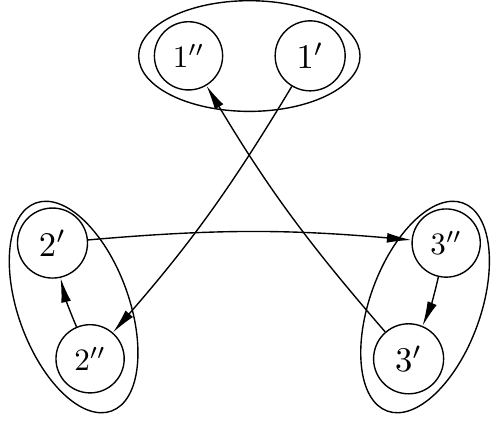}
        \hfil \\ \hline

        \resizebox{.15\textwidth}{!}{
        \begin{tikzpicture}[
        vertex/.style={draw,circle,very thick,minimum size=.6cm},
        arc/.style={draw,very thick,-{Latex[length=3mm, width=2mm]},bend left=10}]
        \node[vertex] (p1) at (0,2.18) {};
        \node[vertex] (p2) at (-1.18,0) {};
        \node[vertex] (p3) at (1.18,0) {};   
        \foreach [count=\r] \row in 
        {{0,1,2},
         {2,0,1},
         {1,2,0}}
        {
            \foreach [count=\c] \cell in \row
            {
                \ifnum\cell=1
                    \draw[arc] (p\r) edge (p\c);
                \fi
                \ifnum\cell=2
                    \draw[arc,cyan] (p\r) edge (p\c);
                \fi
            }
        }
        \end{tikzpicture}
        } &
        \vspace{2mm}

        \includegraphics[width=.18\textwidth]{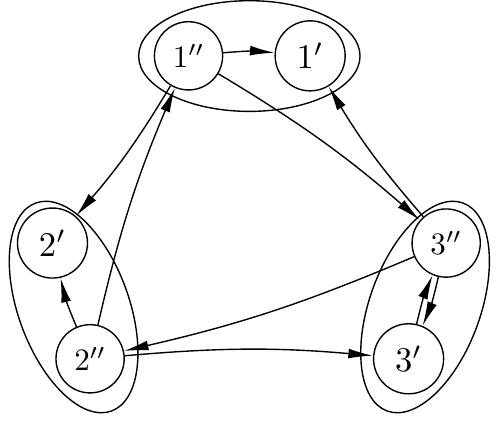}
        \includegraphics[width=.18\textwidth]{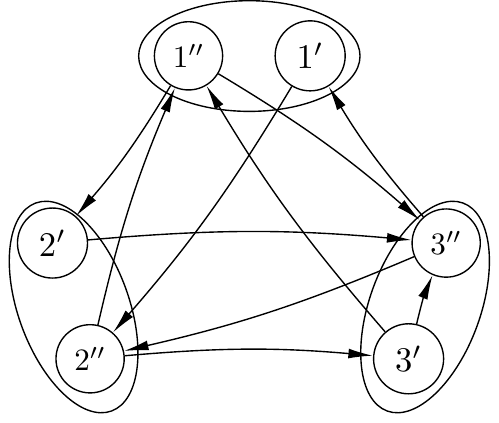}
        \includegraphics[width=.18\textwidth]{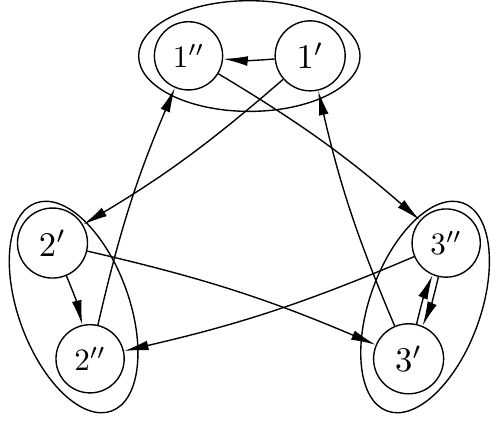}
        \includegraphics[width=.18\textwidth]{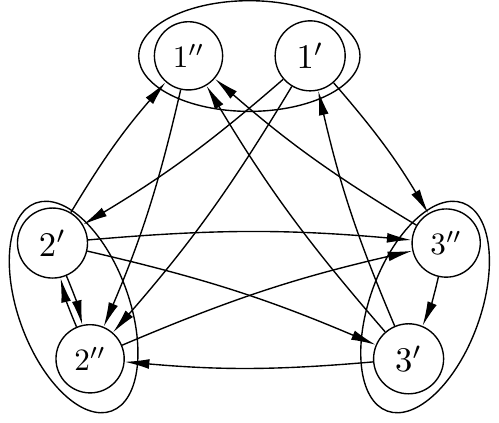}

        \includegraphics[width=.18\textwidth]{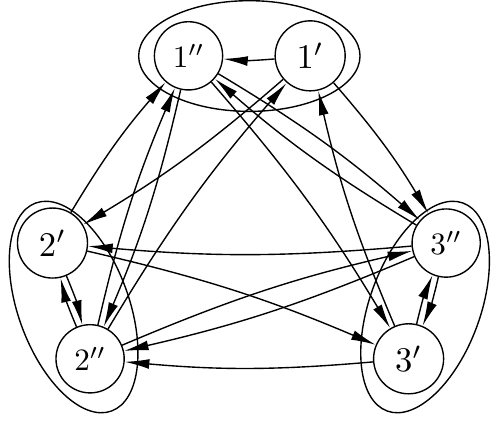}
        \includegraphics[width=.18\textwidth]{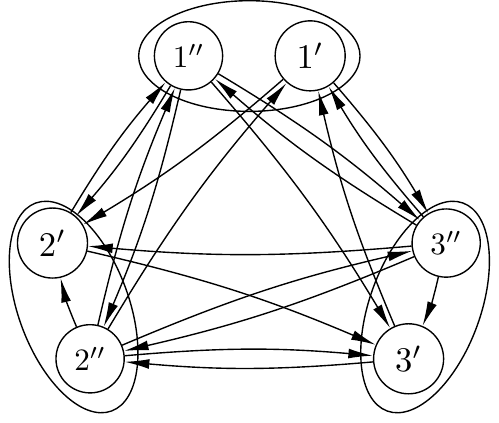}
        \includegraphics[width=.18\textwidth]{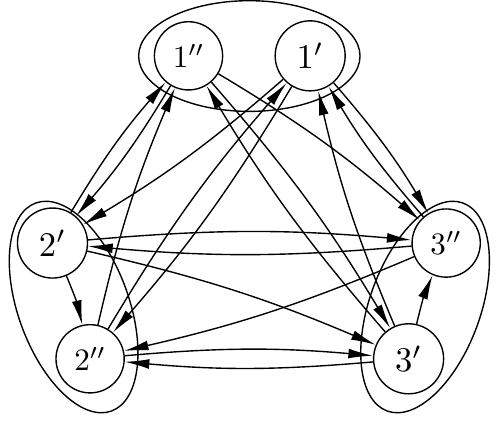}
        \includegraphics[width=.18\textwidth]{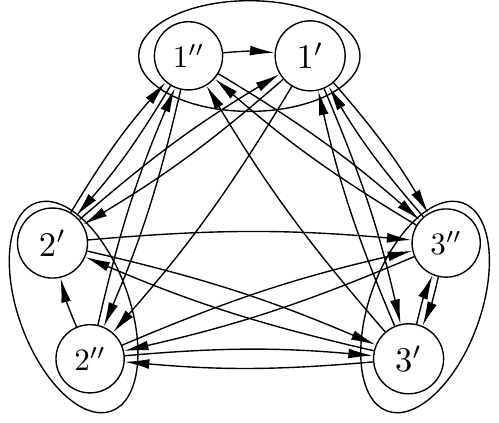} \\
        \hline
    \end{tabular}}
\label{tbl:tbls2}
\end{center}
\end{table*}

\begin{table*}[ht]
\begin{center}
\caption{The $14$ optimal heterogeneous systems with $N=4$ and $L=2$.}
{\renewcommand{\arraystretch}{1.2}
\begin{tabular}{ | M{.16\textwidth} | m{.8\textwidth} | }

        \hline
        symmetric network & \hspace{2in}optimal heterogeneous systems \\
        \hline
        \resizebox{.15\textwidth}{!}{
        \begin{tikzpicture}[
        vertex/.style={draw,circle,very thick,minimum size=.6cm},
        arc/.style={draw,very thick,-{Latex[length=3mm, width=2mm]}}]
        \node[vertex] (p1) at (1,1) {};
        \node[vertex] (p2) at (1,-1) {};
        \node[vertex] (p3) at (-1,-1) {};
        \node[vertex] (p4) at (-1,1) {};    
        \foreach [count=\r] \row in 
        {{0,-1,2,1},
         {1,0,-1,2},
         {2,1,0,-1},
         {-1,2,1,0}}
        {
            \foreach [count=\c] \cell in \row
            {
                \ifnum\cell=-1
                    \draw[arc,red,bend left=10] (p\r) edge (p\c);
                \fi
            }
        }
        \end{tikzpicture}
        } & 

        \includegraphics[width=.18\textwidth]{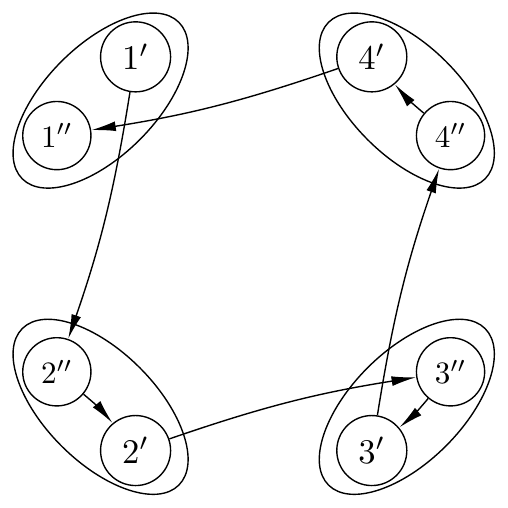}
        \hfil \\ \hline

        \resizebox{.15\textwidth}{!}{
        \begin{tikzpicture}[
        vertex/.style={draw,circle,very thick,minimum size=.6cm},
        arc/.style={draw,very thick,-{Latex[length=3mm, width=2mm]}}]
        \node[vertex] (p1) at (1,1) {};
        \node[vertex] (p2) at (1,-1) {};
        \node[vertex] (p3) at (-1,-1) {};
        \node[vertex] (p4) at (-1,1) {};    
        \foreach [count=\r] \row in 
        {{0,-1,2,1},
         {1,0,-1,2},
         {2,1,0,-1},
         {-1,2,1,0}}
        {
            \foreach [count=\c] \cell in \row
            {
                \ifnum\cell=-1
                    \draw[arc,red,bend left=10] (p\r) edge (p\c);
                \fi
                \ifnum\cell=2
                    \draw[arc,cyan] (p\r) edge (p\c);
                \fi
            }
        }
        \end{tikzpicture}
        } &

        \includegraphics[width=.18\textwidth]{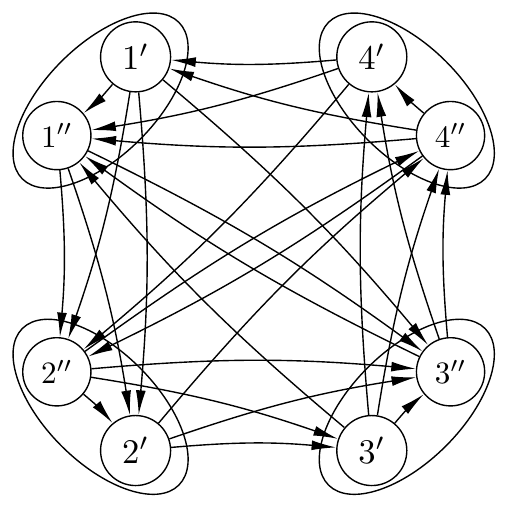}
        \hfil \\ \hline

                \resizebox{.15\textwidth}{!}{
        \begin{tikzpicture}[
        vertex/.style={draw,circle,very thick,minimum size=.6cm},
        arc/.style={draw,very thick,-{Latex[length=3mm, width=2mm]}}]
        \node[vertex] (p1) at (1,1) {};
        \node[vertex] (p2) at (1,-1) {};
        \node[vertex] (p3) at (-1,-1) {};
        \node[vertex] (p4) at (-1,1) {};    
        \foreach [count=\r] \row in 
        {{0,-1,2,1},
         {1,0,-1,2},
         {2,1,0,-1},
         {-1,2,1,0}}
        {
            \foreach [count=\c] \cell in \row
            {
                \ifnum\cell=1
                    \draw[arc,bend left=10] (p\r) edge (p\c);
                \fi
                \ifnum\cell=-1
                    \draw[arc,red,bend left=10] (p\r) edge (p\c);
                \fi
                \ifnum\cell=2
                    \draw[arc,red] (p\r) edge (p\c);
                \fi
            }
        }
        \end{tikzpicture}
        } &

        \includegraphics[width=.18\textwidth]{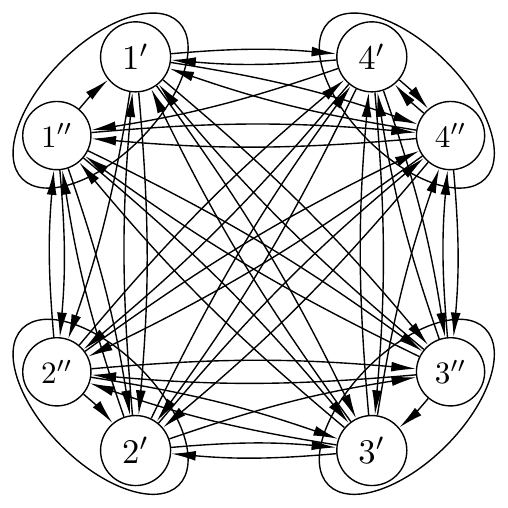}
        \hfil \\ \hline

        \resizebox{.15\textwidth}{!}{
        \begin{tikzpicture}[
        vertex/.style={draw,circle,very thick,minimum size=.6cm},
        arc/.style={draw,very thick,-{Latex[length=3mm, width=2mm]}}]
        \node[vertex] (p1) at (1,1) {};
        \node[vertex] (p2) at (1,-1) {};
        \node[vertex] (p3) at (-1,-1) {};
        \node[vertex] (p4) at (-1,1) {};    
        \foreach [count=\r] \row in 
        {{0,-1,2,1},
         {1,0,-1,2},
         {2,1,0,-1},
         {-1,2,1,0}}
        {
            \foreach [count=\c] \cell in \row
            {
                \ifnum\cell=1
                    \draw[arc,bend left=10] (p\r) edge (p\c);
                \fi
                \ifnum\cell=-1
                    \draw[arc,red,bend left=10] (p\r) edge (p\c);
                \fi
                \ifnum\cell=2
                    \draw[arc,cyan] (p\r) edge (p\c);
                \fi
            }
        }
        \end{tikzpicture}
        } &

        \includegraphics[width=.18\textwidth]{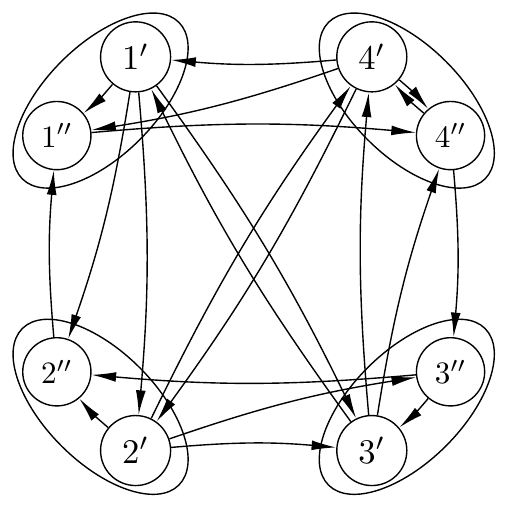}
        \includegraphics[width=.18\textwidth]{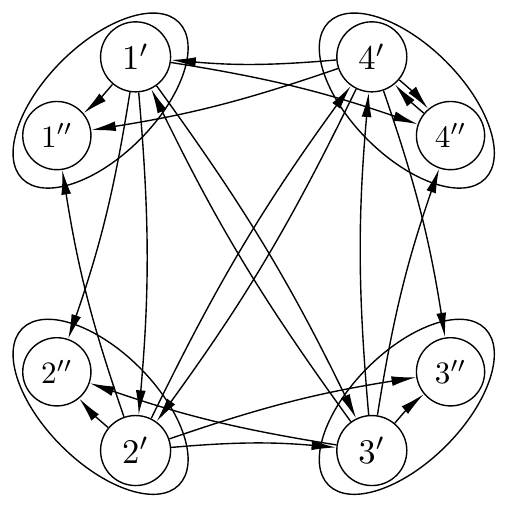}  
        \includegraphics[width=.18\textwidth]{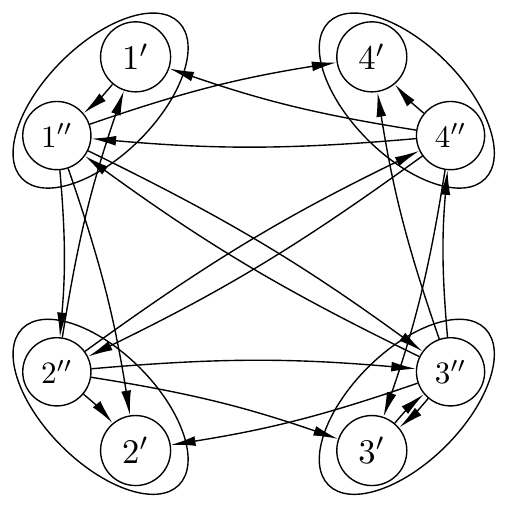}
        \includegraphics[width=.18\textwidth]{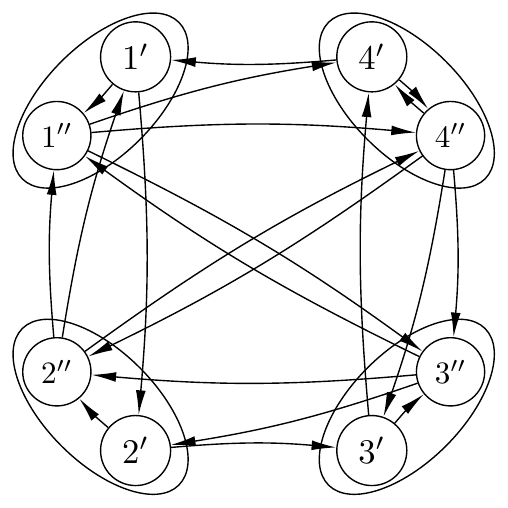}

        \includegraphics[width=.18\textwidth]{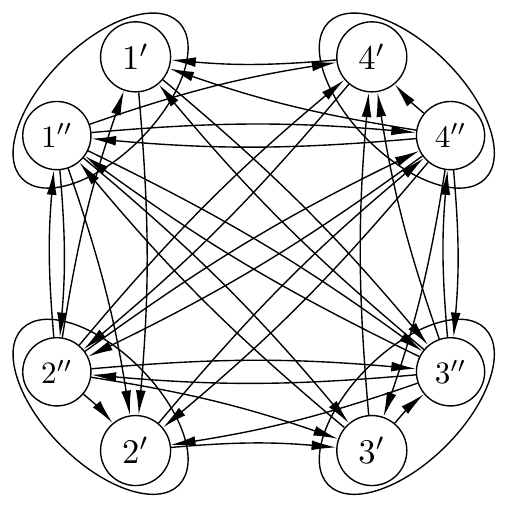}
        \includegraphics[width=.18\textwidth]{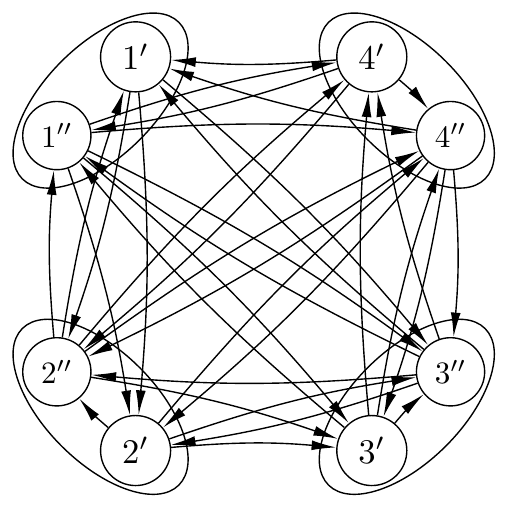}
        \includegraphics[width=.18\textwidth]{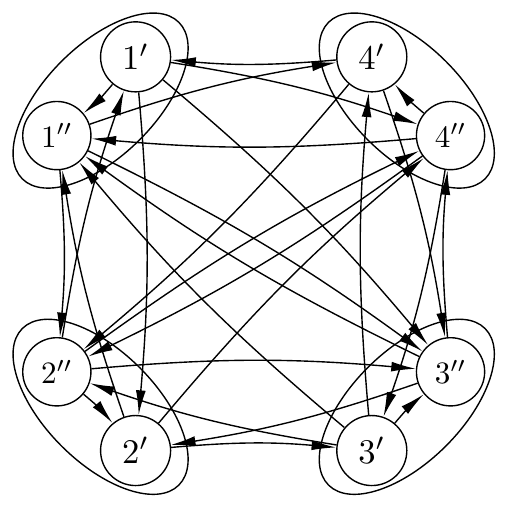}
        \includegraphics[width=.18\textwidth]{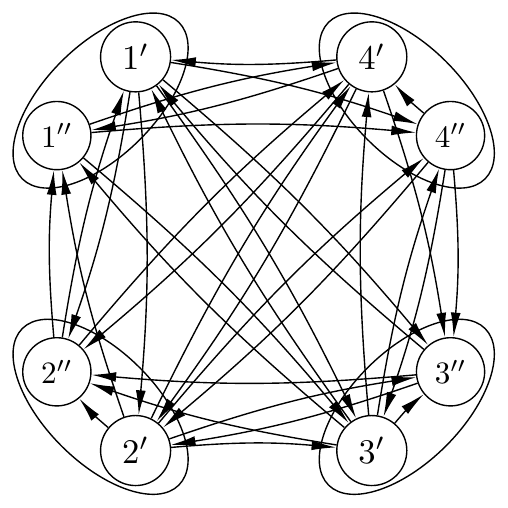}

        \includegraphics[width=.18\textwidth]{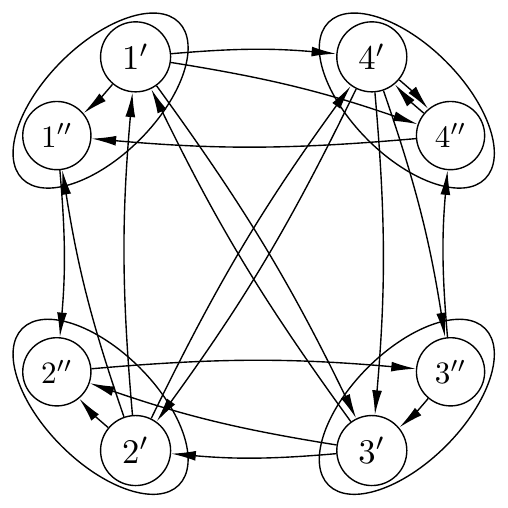}
        \includegraphics[width=.18\textwidth]{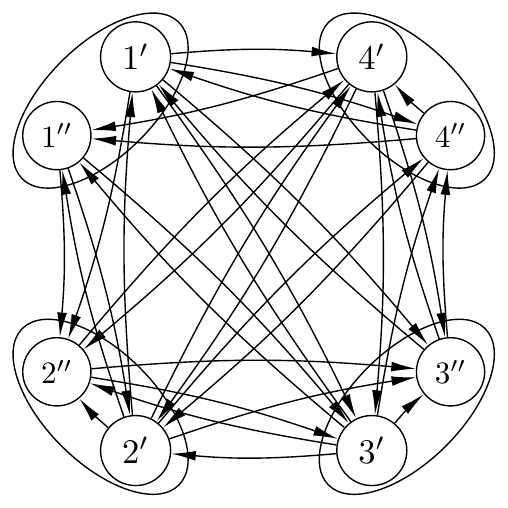} 
        \includegraphics[width=.18\textwidth]{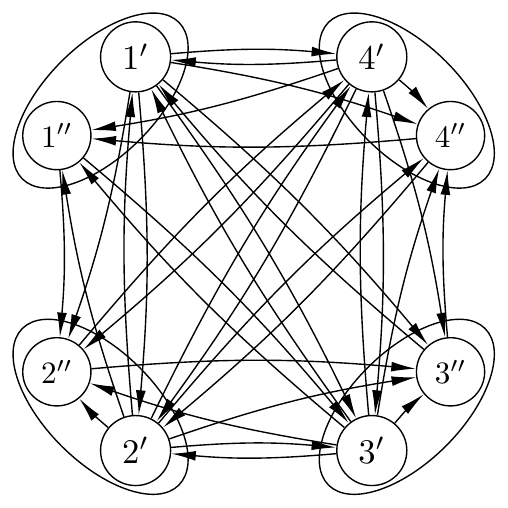} \\
        \hline

    \end{tabular}}
\label{tbl:tbls3}
\end{center}
\end{table*}

\end{document}